\documentclass[aps,prl,floats,floatfix,twocolumn]{revtex4-1}

\usepackage{graphicx}
\usepackage{epstopdf}
\usepackage[caption=false]{subfig}
\graphicspath{{./Figs/}{./figs/}}
\usepackage{float}
\usepackage{latexsym}
\usepackage{amsmath}
\usepackage{amssymb}
\usepackage{bm}
\usepackage{wasysym}
\usepackage{microtype}

\usepackage{color}
\usepackage[colorlinks,allcolors=blue,bookmarks=true]{hyperref}
\usepackage{flafter}
\usepackage[percent]{overpic}
\usepackage{tikz}
\usetikzlibrary{3d}
\usetikzlibrary{positioning}
\usepackage[normalem]{ulem}

\newcommand{\be}[1]{\begin{eqnarray}{\label{e#1}}} 
\newcommand{\beq}{\begin{eqnarray}}
\newcommand{\eeq}{\end{eqnarray}} 

\newcommand{\hide}[1]{}
\newcommand{\Eq}[1]{\textcolor{blue}{{Eq.}\!\!~(\ref{#1})}}

\definecolor{myred}{rgb}  {0.5,0.0,0.0}

\newcommand{\sect}[1]{{\bf #1.}}

\newcommand{\Cn}[1]{\begin{center} #1 \end{center}}

\usepackage{xcolor}

\begin{document}


\title{Probabilistic Hysteresis in Integrable and Chaotic Isolated Hamiltonian Systems}

\author{Ralf~B\"urkle$^{1}$}
\author{Amichay~Vardi$^{2}$}
\author{Doron~Cohen$^{3}$}
\author{James~R.~Anglin$^{1}$}

\affiliation{$^{1}$\mbox{State Research Center OPTIMAS and Fachbereich Physik,} \mbox{Technische Universit\"at Kaiserslautern,} \mbox{D-67663 Kaiserslautern, Germany}}
\affiliation{$^{2}$\mbox{Department of Chemistry,} \mbox{Ben-Gurion University of the Negev,} \mbox{Beer-Sheva 84105, Israel}}
\affiliation{$^{3}$\mbox{Department of Physics,} \mbox{Ben-Gurion University of the Negev,} \mbox{Beer-Sheva 84105, Israel}}

\begin{abstract}
We propose currently feasible experiments using small, isolated systems of ultracold atoms to investigate the effects of dynamical chaos in the microscopic onset of irreversibility. A control parameter is tuned past a critical value, then back to its initial value; hysteresis appears as a finite probability that the atoms fail to return to their initial state even when the parameter sweep is arbitrarily slow. We show that an episode of chaotic dynamics during part of the sweep time produces distinctive features in the distribution of final states that will be clearly observable in experiments.
\end{abstract}


\maketitle

In macroscopic systems, irreversible evolution involves increase of entropy. As for example in the rapid expansion of a gas behind a moving piston, the system's phase space ensemble spreads among instantaneous energy surfaces, thereby spreading \textit{into} a larger region of phase space. Ergodization due to dynamical chaos is then invoked to say that the ensemble effectively \textit{fills} this larger region at reduced coarse-grained density, so that entropy increases even though Liouville's theorem forbids phase space volume change in isolated systems. 

A striking form of irreversibility can be observed in hysteresis experiments: A control parameter is slowly changed and then changed back to its initial value, and yet the system does not return to its initial state. Recent experiments have studied hysteresis in dissipative open systems \cite{Eckel, Trenkwalder}. In this work we discover a new mechanism for irreversibility in \textit{isolated} systems, which we call Hamiltonian Hysteresis.  This mechanism is related to topological structure in phase space.

The standard mechanism for irreversibility, as in the piston paradigm, is based on linear response (Kubo) theory for dissipation. For an isolated system, this theory requires the assumption of \textit{chaos}, as outlined by Ott, Wilkinson and followers \cite{Ott1,Ott2,Ott3,Wilkinson1,Wilkinson2,Cohen}. 
The standard theory forbids quasi-static entropy growth, because linear response, by definition, provides zero energy spreading in the adiabatic limit. In contrast, Hamiltonian Hysteresis can provide irreversibility even in the quasi-static limit.

Hamiltonian Hysteresis can occur probabilistically even without chaos \cite{dimer}. Chaos does however greatly enhance irreversibility and has observable signatures, as we demonstrate in a small isolated system.
 We propose to extend recently suggested experiments \cite{dimer} to show these fingerprints of chaos in irreversibility. In particular we 
{\bf (1)} explain how irreversibility can arise even in quasi-integrable dynamics; 
{\bf (2)} identify fingerprints of \textit{quasi-static passage through chaos}; 
and {\bf (3)} distinguish these from the effects of Kubo-Ott energy spreading for non-zero sweep rates.

\sect{Hamiltonian hysteresis}  
A simple example of Hamiltonian hysteresis is provided by a classical particle in a double well, where the relative depths of the two wells are slowly tuned over time. If the particle is initially orbiting within the lowest well, it stays within this well adiabatically until the slowly time-dependent well becomes so shallow that the particle escapes over the inter-well potential barrier. When there is no dissipation from a macroscopic reservoir to drag the particle down to the bottom of the second well (as for example in \cite{Trenkwalder}), the particle continues flying above the barrier, back and forth across both wells. If the relative depths of the two wells are then slowly returned to their initial values, however, there is a subtlety in adiabatic mechanics involving the breakdown of adiabaticity around a separatrix even for arbitrarily slow change of parameters \cite{Timofeev,Tennyson,Hannay,Cary,Elskens}. Whether the particle ends up orbiting within the original well after this, or is instead found in the other well, turns out to depend sensitively on the phase of the particle in its initial orbit as well as on the timing of the potential change. If these are not both controlled to high precision, it will be probabilistic whether the particle ends up back in its initial state or in a dramatically different one \cite{Kruskal,Neishtadt1,Henrard,Neishtadt2,LC,dimer}.

A system similar to the double well but which can include quantum many-body effects and also be realized experimentally is the two-mode Bose-Hubbard model (`dimer'). In \cite{dimer} we showed how probabilistic hysteresis in such an integrable Hamiltonian system can be described quantitatively in terms of expanding and filling phase space volumes, just as in the usual statistical mechanical theory, showing that this simple hysteresis is truly a microscopic limit of macroscopic irreversibility. To investigate the role of dynamical chaos in the microscopic onset of irreversibility, we now turn to a class of realizable model systems which can be tuned to be either integrable or chaotic: the $3$-site Bose-Hubbard `trimer'.

\sect{Proposed experiments} 
Our testing ground system consists of $N$ condensed bosons in an optical lattice with $3$ sites. In the tight-binding Bose-Hubbard limit, its Hamiltonian is
\begin{equation} 
\hat H=\frac{x(t)}{2}(\hat n_1-\hat n_3)+\frac{U}{2}\sum_{j=1}^3 \hat n_j^2-\frac{\Omega}{2} \sum_{j=1}^{2}\left(\hat a^{\dagger}_{j+1}\hat a_j+h.c.\right).
\label{eq:H}
\end{equation}
Here $\hat{n}_{j}=\hat{a}^{\dagger}_{j}\hat{a}_{j}$ are the occupations of the $j$-th site, 
$U<0$ is the attractive on-site interaction, and $\Omega$ is the inter-site hopping frequency.  
The external parameter $x(t)$ controls the potential bias. 
Our hysteresis scenario is the forward-and-back sweep
\begin{equation} \label{e2}
x(t) = x_0 + (x_I-x_0) \frac{|t|}{T},  
\ \ \ \ \ \ {-T<t<T}
\end{equation}
The control parameter $x(t)$ is swept from $x_I$ at the initial time $t=-T$ to $x_0$ at $t=0$, and then swept back to $x_I$ at $t=+T$, with sweep rate $\dot{x}\propto 1/T$.
\begin{figure}
\subfloat[]{\includegraphics[width=0.24\textwidth]{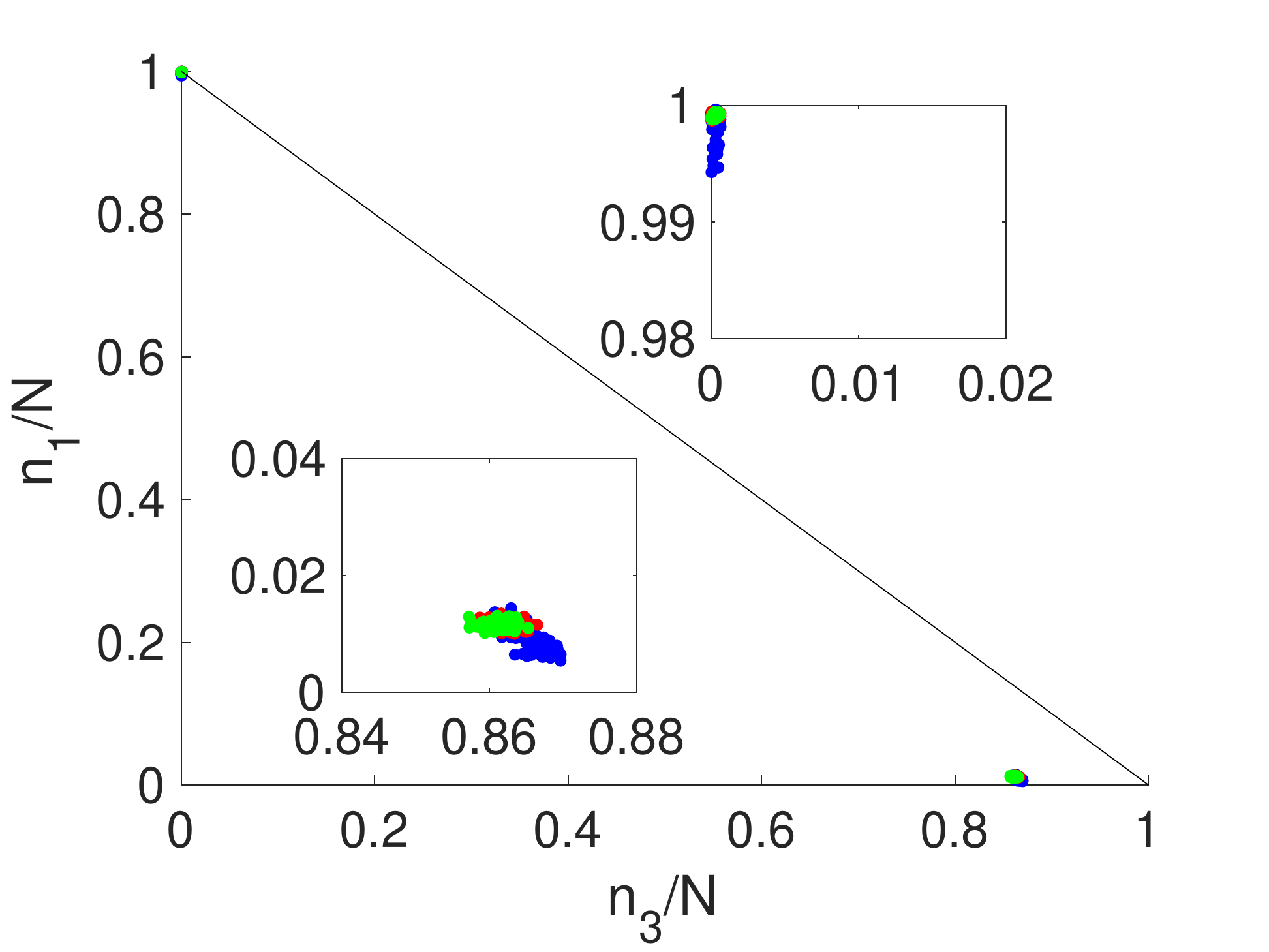}}
\subfloat[]{\includegraphics[width=0.24\textwidth]{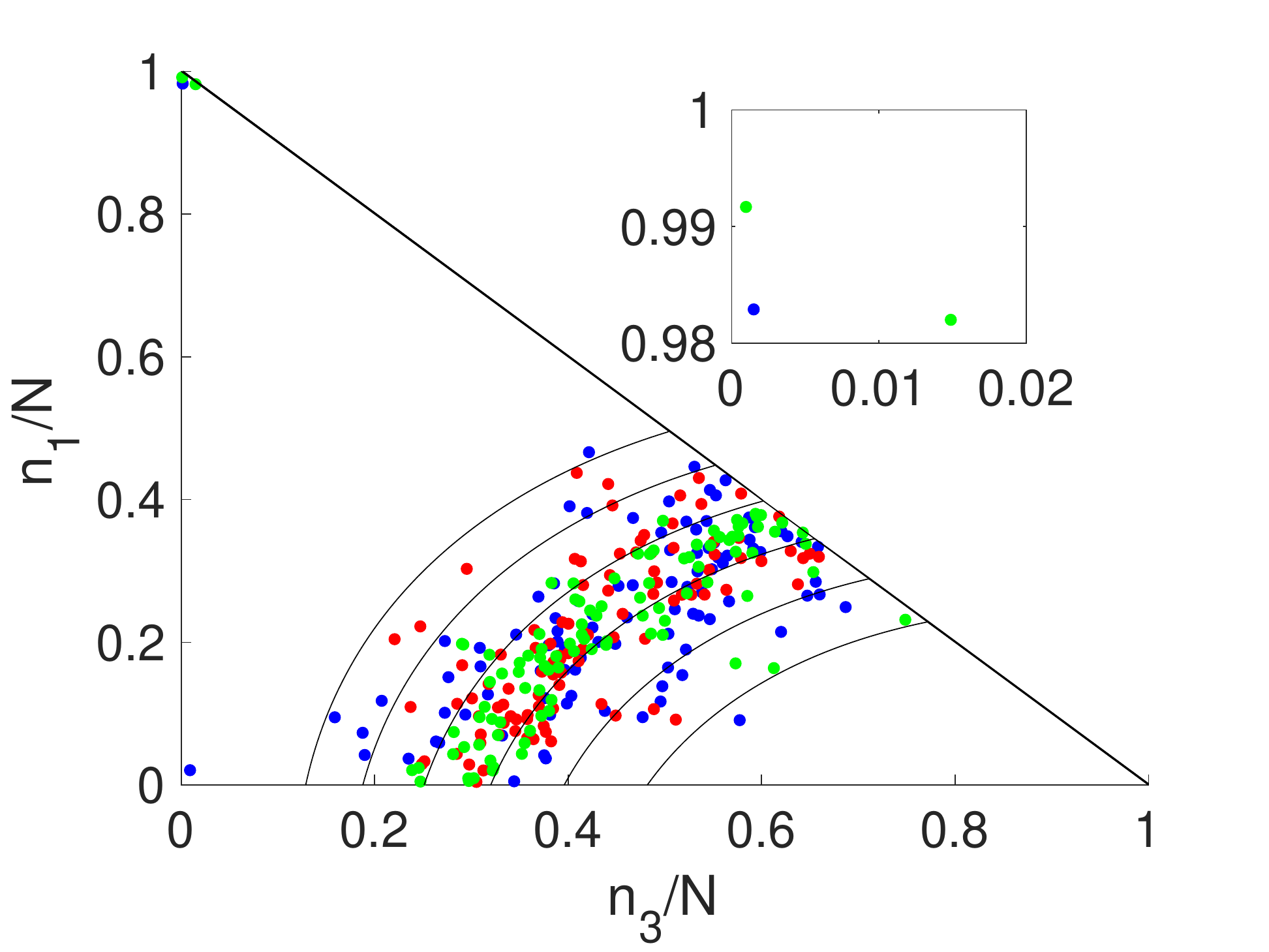}}\\
\vspace{-3mm}
\subfloat[]{\includegraphics[width=0.24\textwidth]{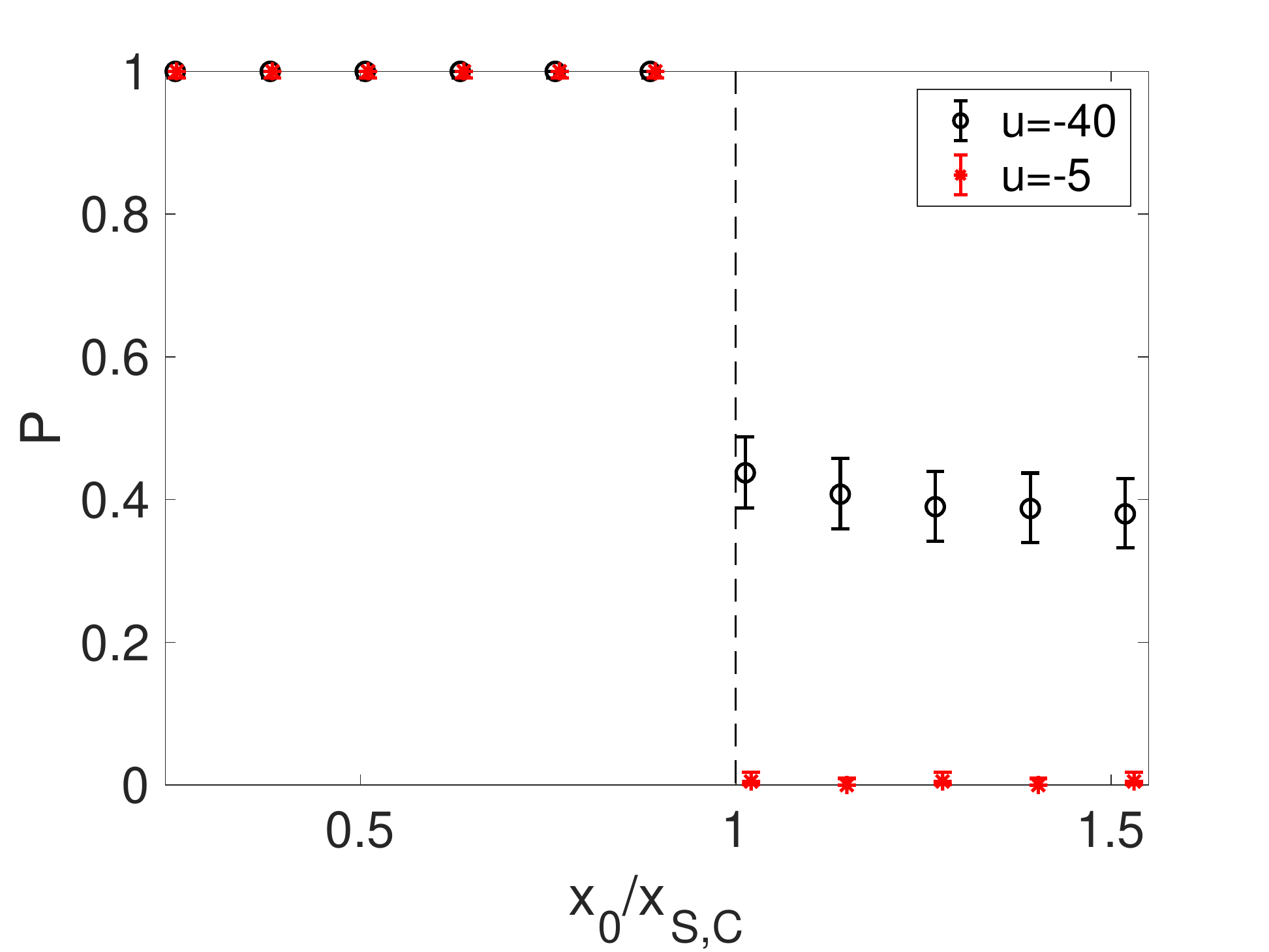}}
\subfloat[]{\includegraphics[width=0.24\textwidth]{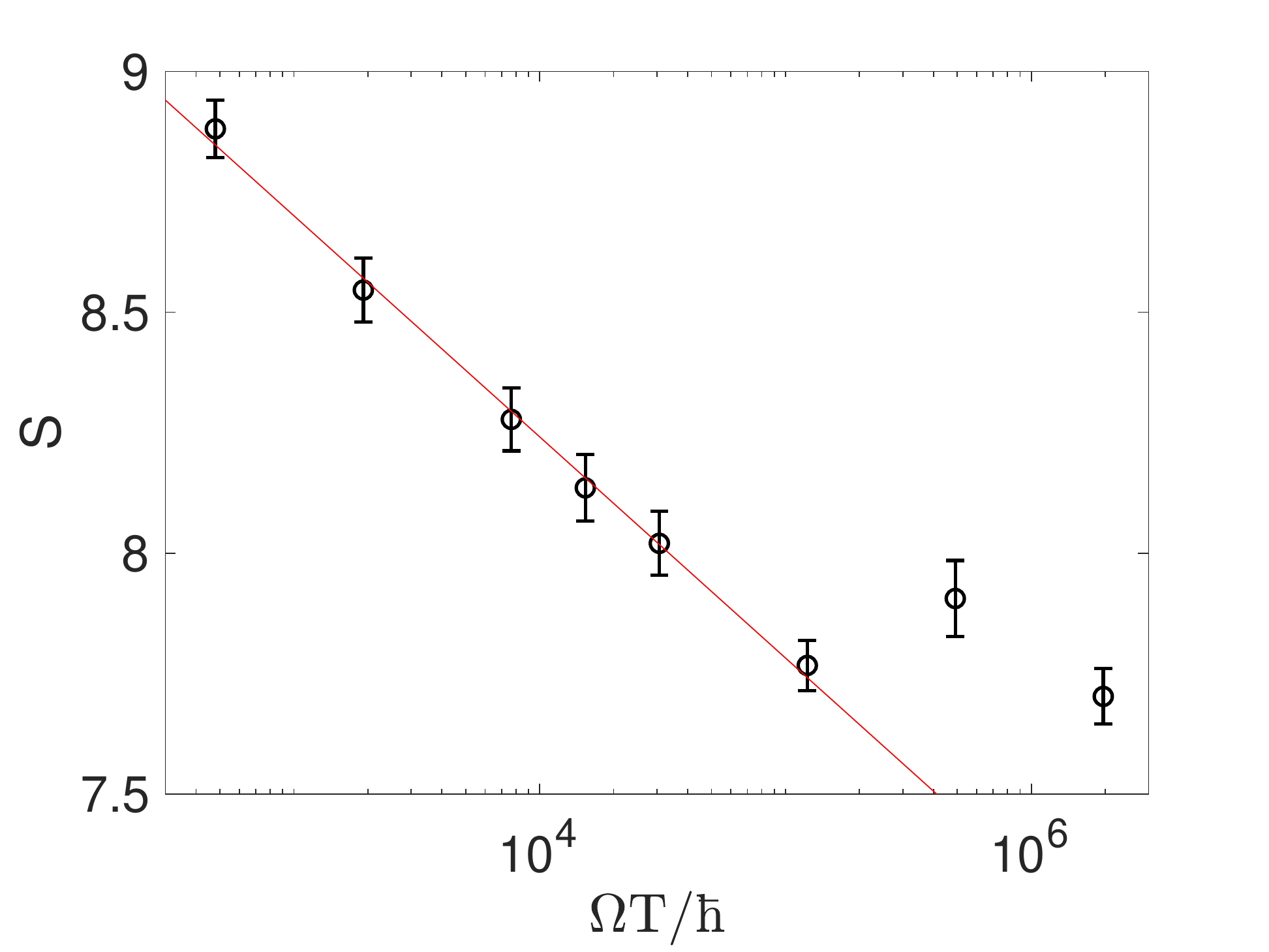}}
\caption{\label{f1}
Simulations of Hamiltonian hysteresis with \Eq{eq:H}.
The occupations $(n_{1},n_{3})$ at the end of the sweep protocol \Eq{e2}.  
In the upper panels each dot represents a separate run of the experiment. Insets zoom into the relevant regions (upper left and lower right corner) to show the time dependence of the size of the cloud in (a) and the very low return probability in (b).
Different colors correspond to different sweep rates:
\textbf{(a)}~For $u=-40$, with $x_{I}/\Omega=-100$ and $x_0/\Omega=60$ (blue: $\Omega T/\hbar=5333$, red: $\Omega T/\hbar=53333$, green: $\Omega T/\hbar=106667$.).  
\textbf{(b)}~For $u=-5$, with $x_I/\Omega=-40$ and $x_{0}/\Omega=6$ (blue: $\Omega T/\hbar=7667$, red: $\Omega T/\hbar=30667$, green: $\Omega T/\hbar=122667$. The black curves are contours of constant energy just after the exit from chaos (for details see \cite{SMe}). Different contours correspond to different energies.)
\textbf{(c)}~Return probabilities $P(x_{0})$ defined as the fraction of 400 runs that comes back to the initial distribution. 
\textbf{(d)}~Dependence of the entropy $S$ (log area of the $n$-distribution) on the sweep time $T$ for simulations in case~(b). The red line shows a fit to $-a \log(\Omega T/\hbar)+S_0$ before $S$ levels off to a residual value. For the simulation in (a) the width of the distribution is negligible.}
\end{figure}

We assume negative $x_I$ large enough that for early and late times hopping between sites is adiabatically suppressed (they are far detuned). During these initial and final stages, therefore, the three occupation numbers $\hat{n}_j$ are adiabatically invariant and evolution is trivial. With large $-x_I$ we assume that an initial state can be prepared at $t=-T$ which is cold enough that most atoms are in the lowest-energy site $j=1$. We do \textit{not} require an initial coherent state of definite phase, but a low-temperature thermal mixed state. To simplify our explanation of the subsequent evolution we idealize this realistic state as a narrow, low energy microcanonical state. A more realistic canonical distribution will show similar behavior, just with a larger range of energies. The data which will show hysteresis will be the occupation numbers $\hat{n}_j$, measured at $t=+T$ and recorded for many runs (repeated or parallel) of the experiment. High-precision atom counting is \textit{not} required because the final states which must be distinguished differ by significant numbers. Both state preparation and final read-out are performed at times when the system dynamics is trivial, and so both are cleanly separated as physical processes from the nontrivial dynamics which they allow us to see.

\sect{Simulated Results}
Throughout this paper we describe the quantum many-body evolution of our Bose-Hubbard system in the semiclassical truncated Wigner approximation, evolving an initial ensemble representing the quantum mixed state with the Gross-Pitaevskii equation (discrete non-linear Schr\"odinger equation) that is associated with \Eq{eq:H} \cite{SMa}. This approximation should be accurate for attainably large particle numbers $N$.

Numerical results simulating Hamiltonian hysteresis for two representative values of the interaction parameter $u=UN/\Omega$ are shown in Fig.~\ref{f1}. Measurements of $n_{1,3}$ in the initial state at $t=-T$ will show a distribution like that shown in panels~1a and~1b: almost all atoms are in site 1. If $x_0$ is below a critical value $x_S$ when $u=-40$, or below a threshold $x_C$ when $u=-5$, the system returns to its initial state at $t=+T$. However, if the parameter sweep extends beyond these thresholds, some experimental runs end up in significantly different states with different $n_{1,3}$. We define the fraction of runs in which the final $n_{1,3}$ populations are indistinguishable from their initial values to be the measured \textit{return probability} $P(x_{0})$. Dramatic differences are observed between the post-threshold final population distributions of Fig.~\ref{f1}(a) and Fig.~\ref{f1}(b). Whereas the final $n_{1,3}$ values for $u=-40$ are nearly binary, \textit{i.e.} tightly localized around either the original values or around a single alternative, the $n_{1,3}$ ensemble for $u=-5$ traces a strip whose width becomes smaller as the sweep becomes slower. Rather than vanish in the limit of  an infinitely slow sweep, the strip's width saturates to a finite value implying a significant final-state entropy.

In what follows we show that these observations provide a clear and experimentally observable hallmark of chaos in Hamiltonian hysteresis. For $u=-40$, the classical dynamics of \Eq{eq:H} remains integrable throughout our hysteresis scenario. Irreversibility is then generated by essentially the same mechanism that produces it in the double-well example above, or in the Bose-Hubbard dimer~\cite{dimer}. Here it is the simple merging, at $x=x_S$, of two tori in phase space \cite{SMb,SMc}. By contrast, for $u=-5$, the classical dynamics becomes chaotic at $x>x_C$, with dramatic effects on the final $n_j$ distribution \cite{SMc}. 

\sect{Phase space structure} 
Discounting the conserved total particle number $N$, the Bose-Hubbard trimer has two degrees of freedom, with two pairs of canonical variables $(q_1,q_2,p_1,p_2)$ spanning its phase space \cite{SMa}. Energy surfaces are therefore 3D and the dynamics in any given region of phase space can be either quasi-integrable or chaotic, depending on the values of $u$ and $x$. In the integrable regions of phase space, we can define action angle variables $(I_a,\varphi_a)$ and $(I_b,\varphi_b)$, so that the quasi-static motion at any given value of $x$ takes place on the surface of the 2D tori defined by the actions $I_{a,b}$. Each 3D energy surface consists of multiple tori, all satisfying $H(I_a,I_b;x)=E$. 

\sect{Different types of adiabaticity} 
During periods of integrable motion, adiabaticity is maintained in the Einstein-Landau sense \cite{Landau}, as the conservation of $I_{a,b}$ under sufficiently slow variation of $x$. In our scenario this means that $1/T \ll \omega_B$, where $\omega_B$ are the perturbative Bogoliubov frequencies around the followed stationary point \cite{SMb}. The system remains on a single $\bm{I}\equiv(I_a,I_b)$ torus whose energy changes as $E(t)=H[\bm{I};x(t)]$. 
By contrast, when the dynamics becomes chaotic, the $I_a$ and $I_b$ motions become coupled and can exchange energy. Rather than follow a single torus, the system is then free to ergodize over the entire 3D energy surface (or over the chaotic part of a mixed energy surface). As long as chaos prevails, the system's energy will follow the adiabatic energy surface, as discussed by Ott \cite{Ott1}. Unlike in integrable adiabaticity where the two actions are adiabatic invariants, the single adiabatic invariant in the Ott regime is the enclosed phase-space volume.

While the actions $I_{a,b}$ are generally rather complex functions of the canonical variables $q_{1,2},p_{1,2}$, their form at $t=\pm T$ is quite simple: they correspond closely to the measured populations $n_{1,3}$. And since the actions do not change during integrable motion, our final population distribution can be considered as a snapshot of the action distribution upon exit from the Ott regime.

\begin{figure*}
\subfloat[]{
\begin{tikzpicture}
\node (a) at (0,0) {\includegraphics[width=.19\textwidth, trim=0mm 0mm 18mm 9mm, clip]{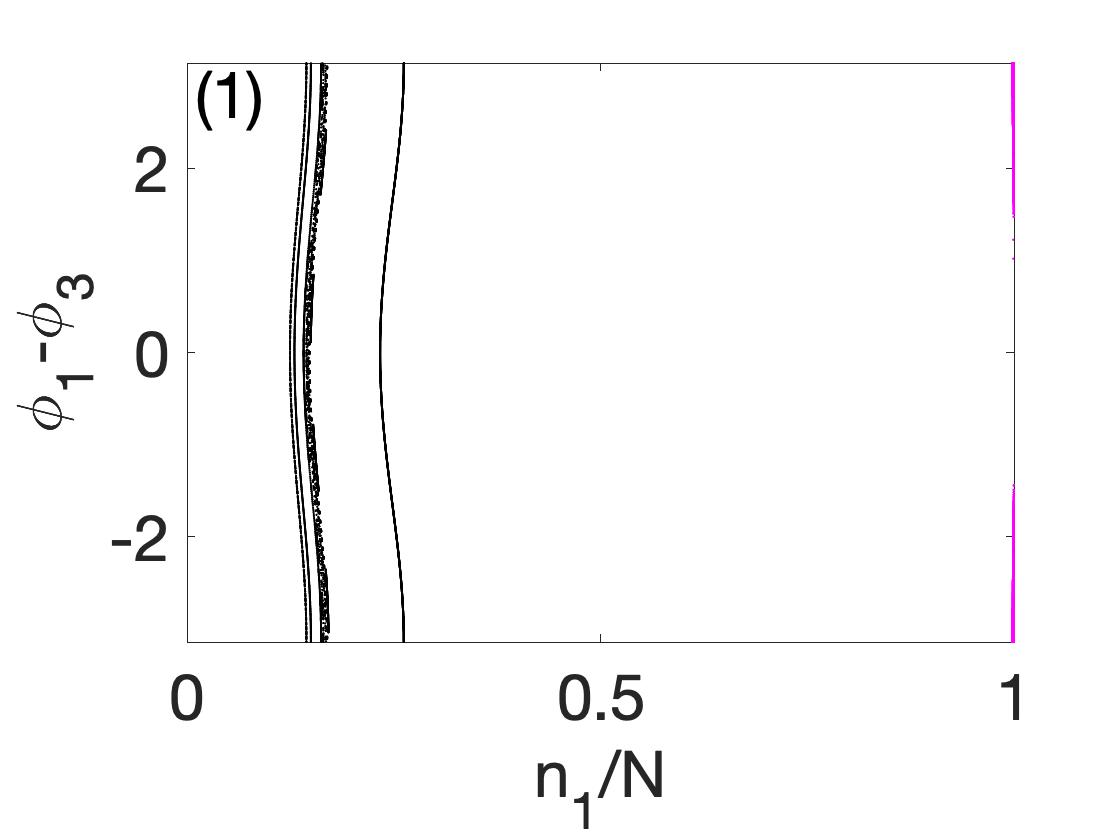}}; 
\node[ right=0mm of a] (b)  {\includegraphics[width=.19\textwidth, trim=25mm 12mm 18mm 9mm, clip]{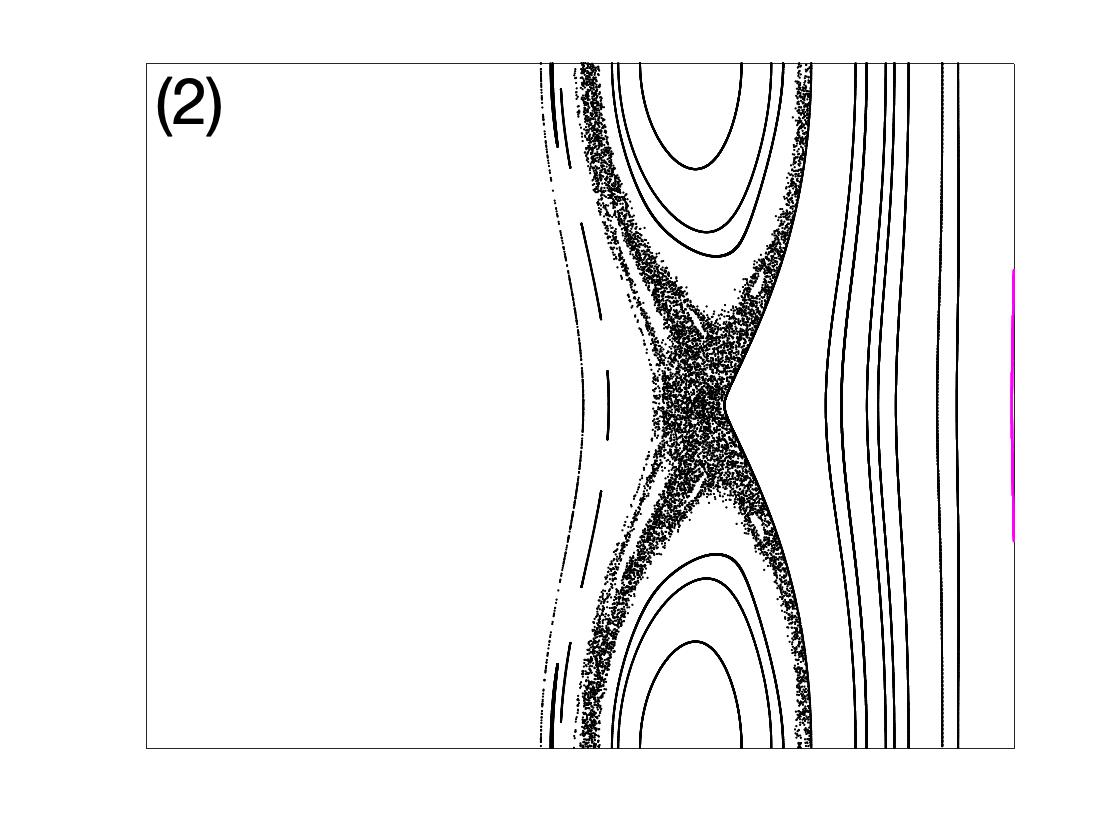}};
\node[right=0mm of b] (c)  {\includegraphics[width=.19\textwidth, trim=25mm 12mm 18mm 9mm, clip]{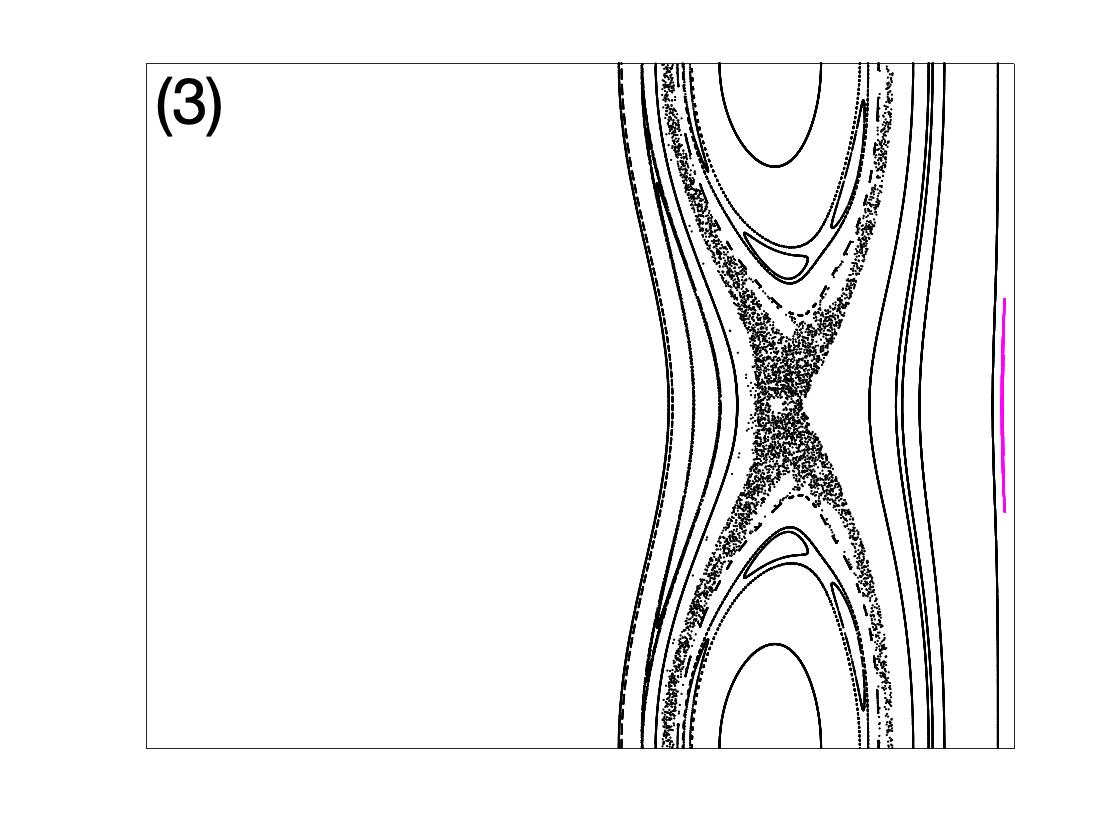}};
\node[right=0mm of c] (e)  {\includegraphics[width=.19\textwidth, trim=25mm 12mm 18mm 9mm, clip]{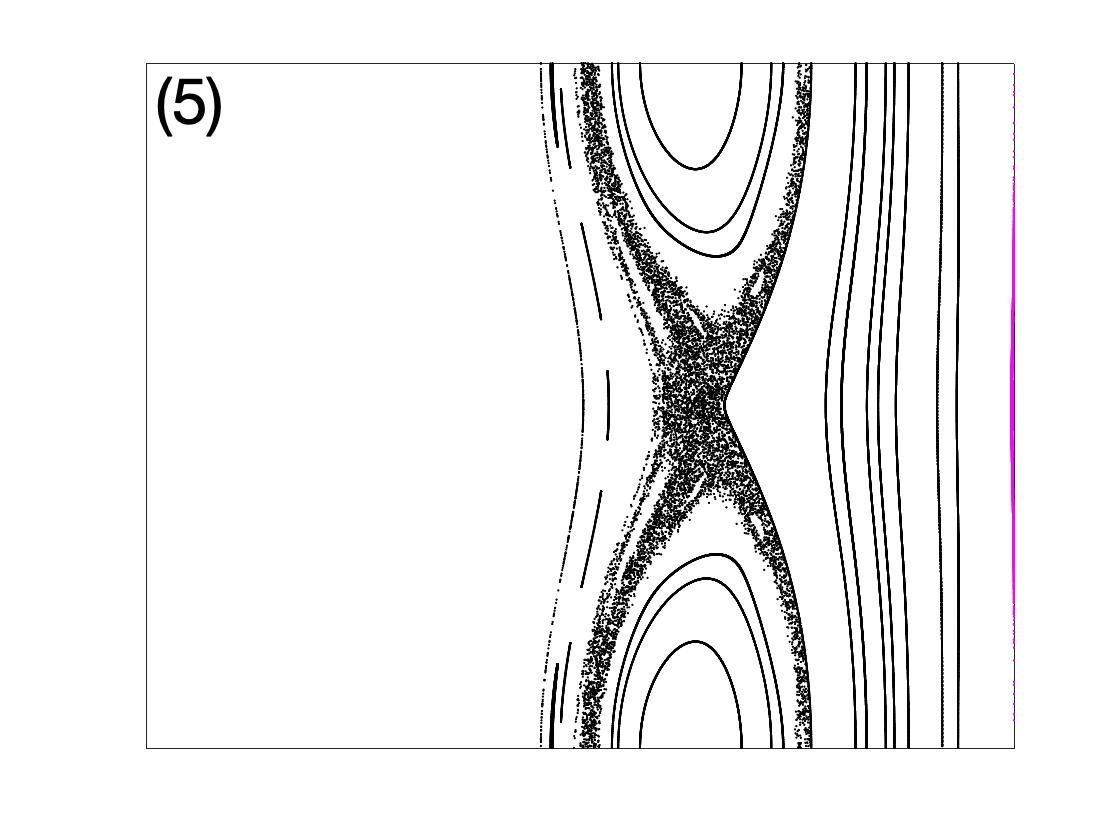}};
\node[right=0mm of e] (g)  {\includegraphics[width=.19\textwidth, trim=25mm 12mm 18mm 9mm, clip]{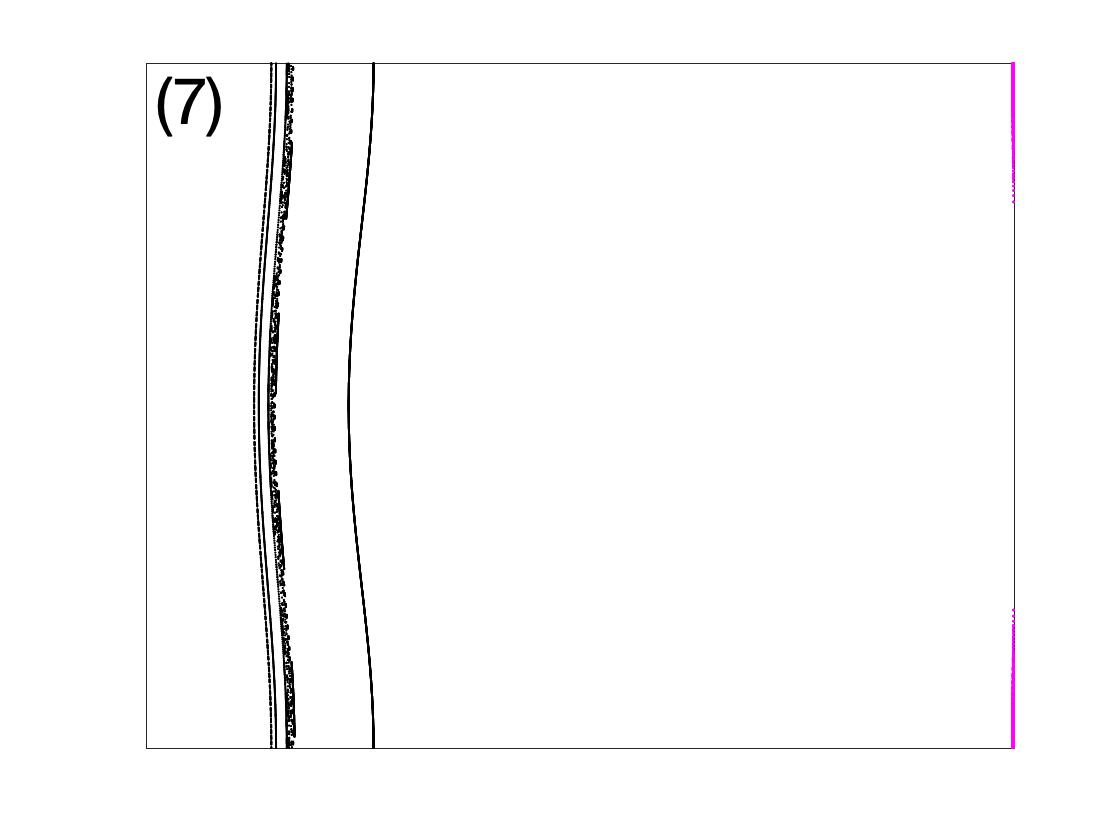}};
\node[above=0mm of e] (d) {\includegraphics[width=.19\textwidth, trim=25mm 12mm 18mm 9mm, clip]{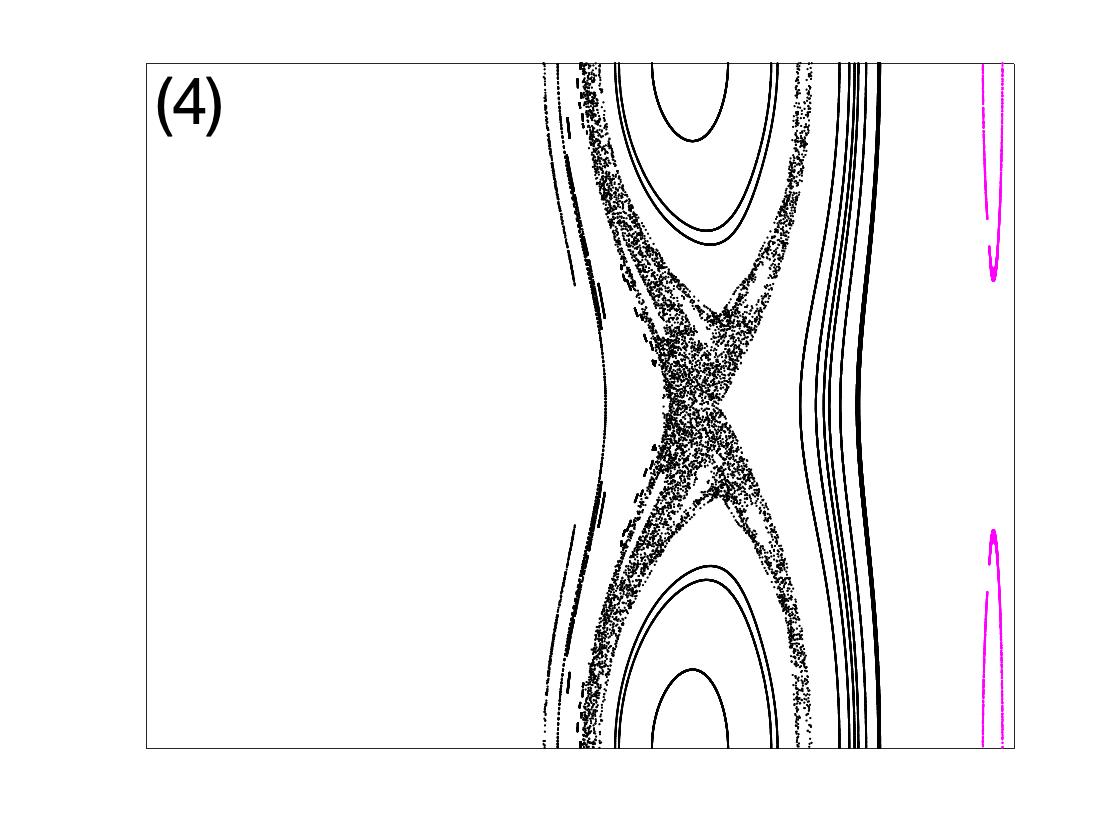}};
\node[right=0mm of d] (f)  {\includegraphics[width=.19\textwidth, trim=25mm 12mm 18mm 9mm, clip]{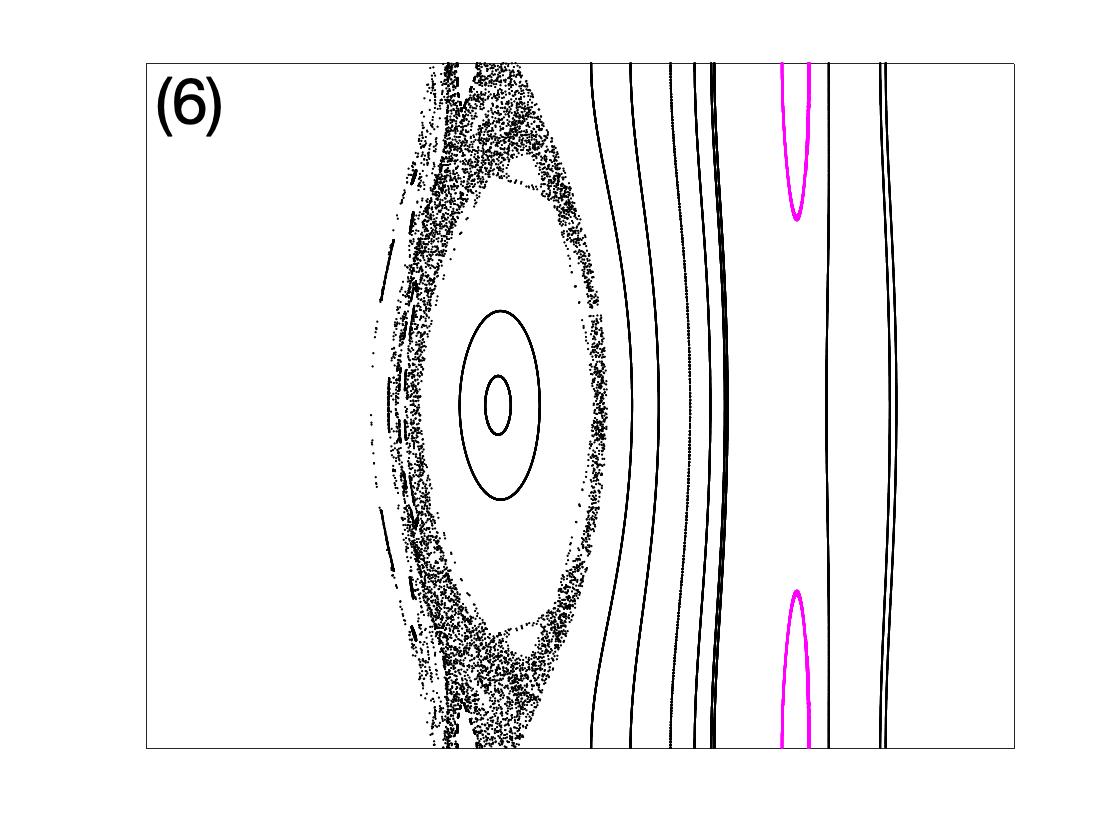}};
\node[above=0mm of b] (h)  {\includegraphics[width=.6\textwidth, trim=5mm 0mm 10mm 7mm, clip]{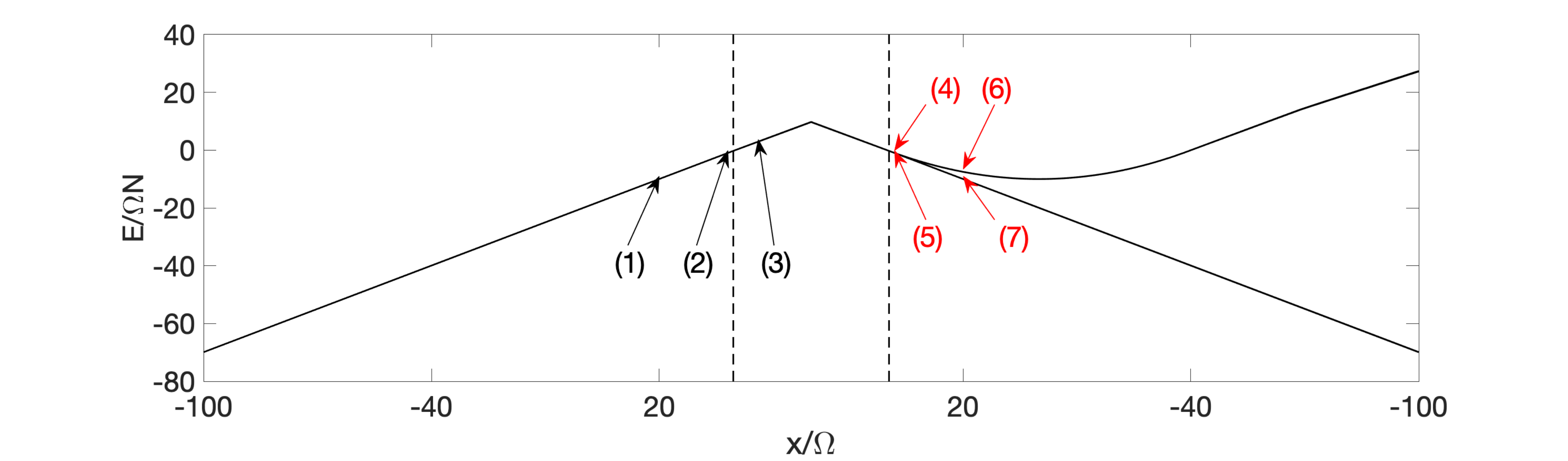}};
\end{tikzpicture}}

\subfloat[]{
\begin{tikzpicture}
\node (a) at (0,0) {\includegraphics[width=.19\textwidth, trim=5mm 2mm 12mm 9mm, clip]{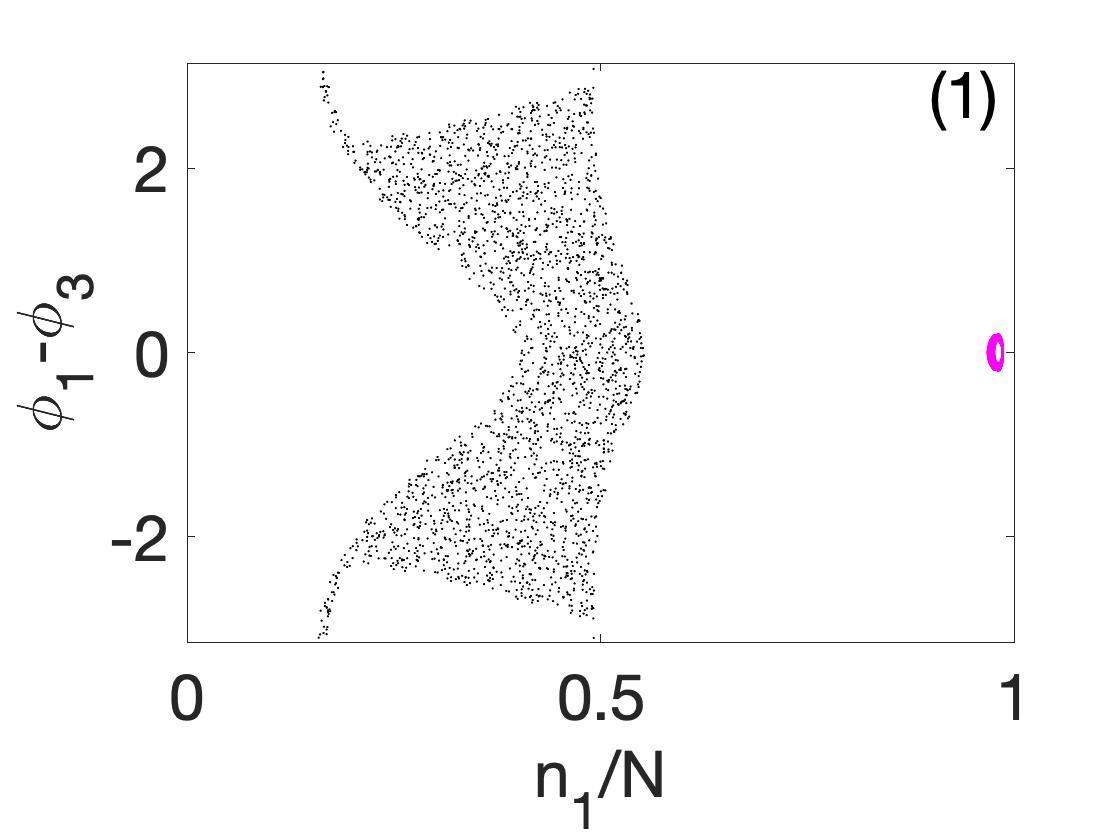}}; 
\node[right=0mm of a] (b)  {\includegraphics[width=.19\textwidth, trim=20mm 2mm 12mm 9mm, clip]{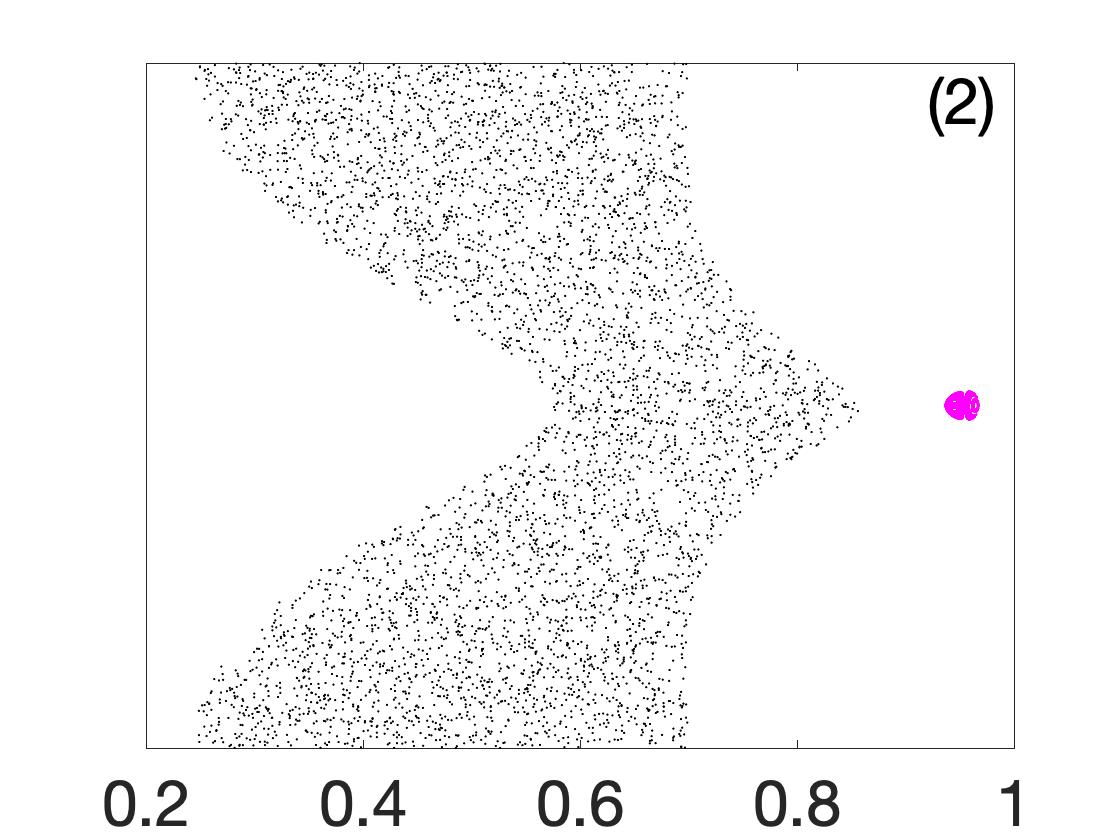}};
\node[ right=0mm of b] (c)  {\includegraphics[width=.19\textwidth, trim=20mm 2mm 12mm 9mm, clip]{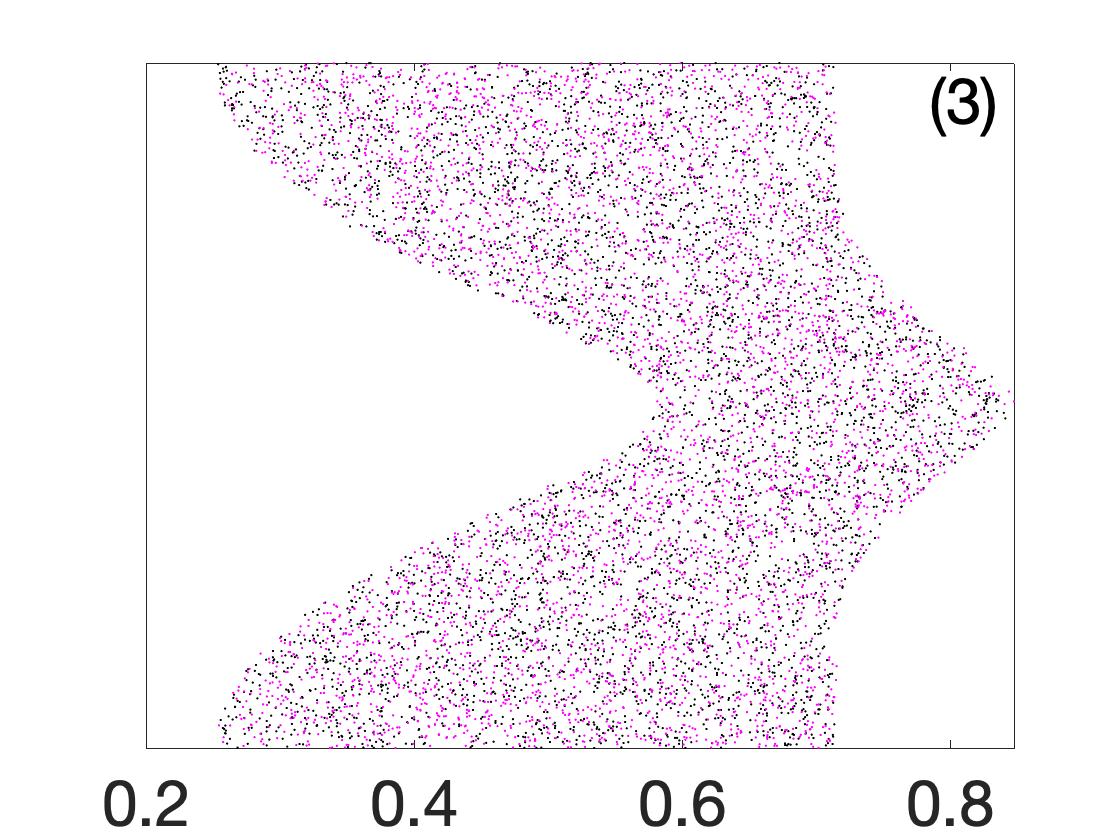}};
\node[right=0mm of c] (f)  {\includegraphics[width=.19\textwidth, trim=20mm 2mm 12mm 9mm, clip]{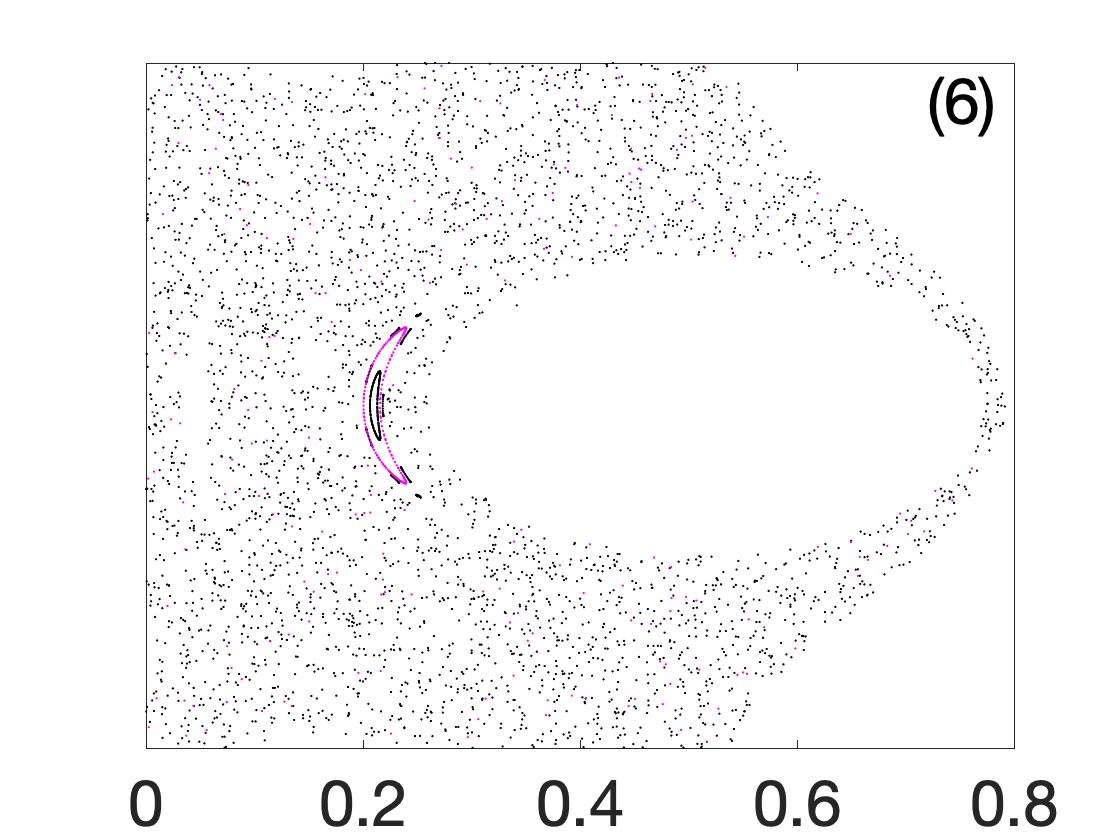}};
\node[above=0mm of f] (e)  {\includegraphics[width=.19\textwidth, trim=20mm 12mm 12mm 9mm, clip]{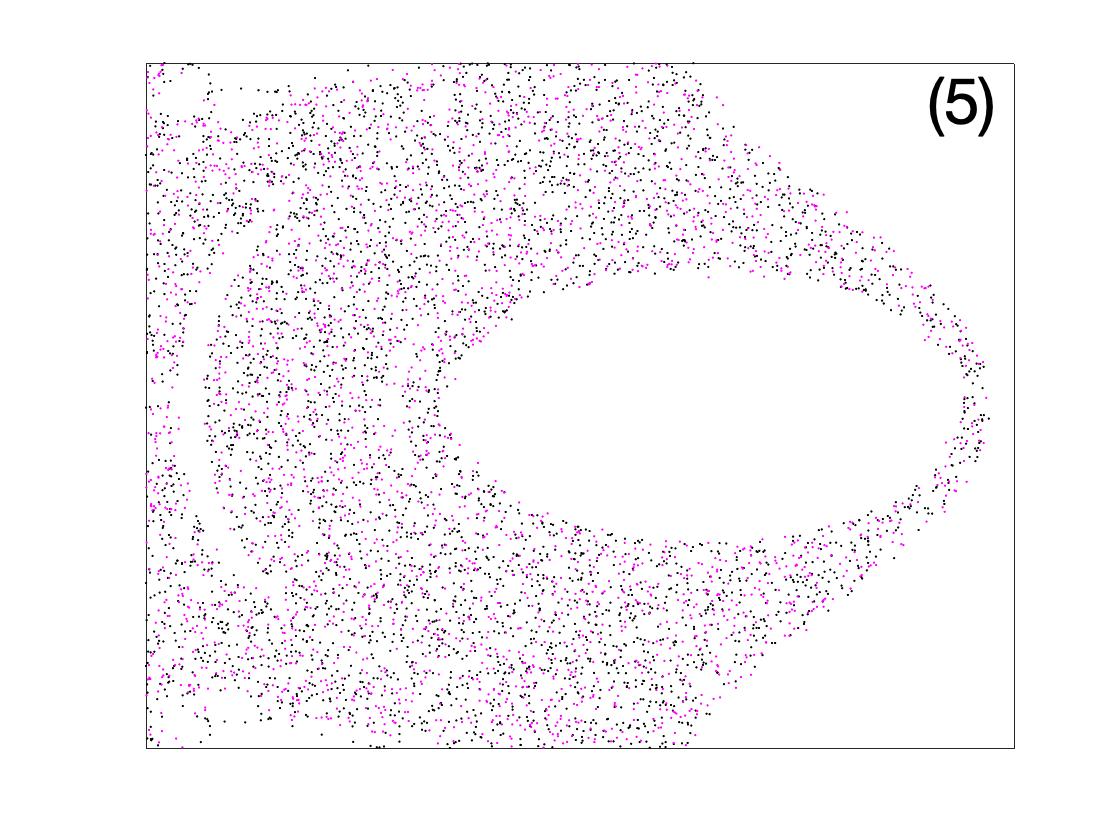}};
\node[above=0mm of e] (d) {\includegraphics[width=.19\textwidth, trim=20mm 12mm 12mm 9mm, clip]{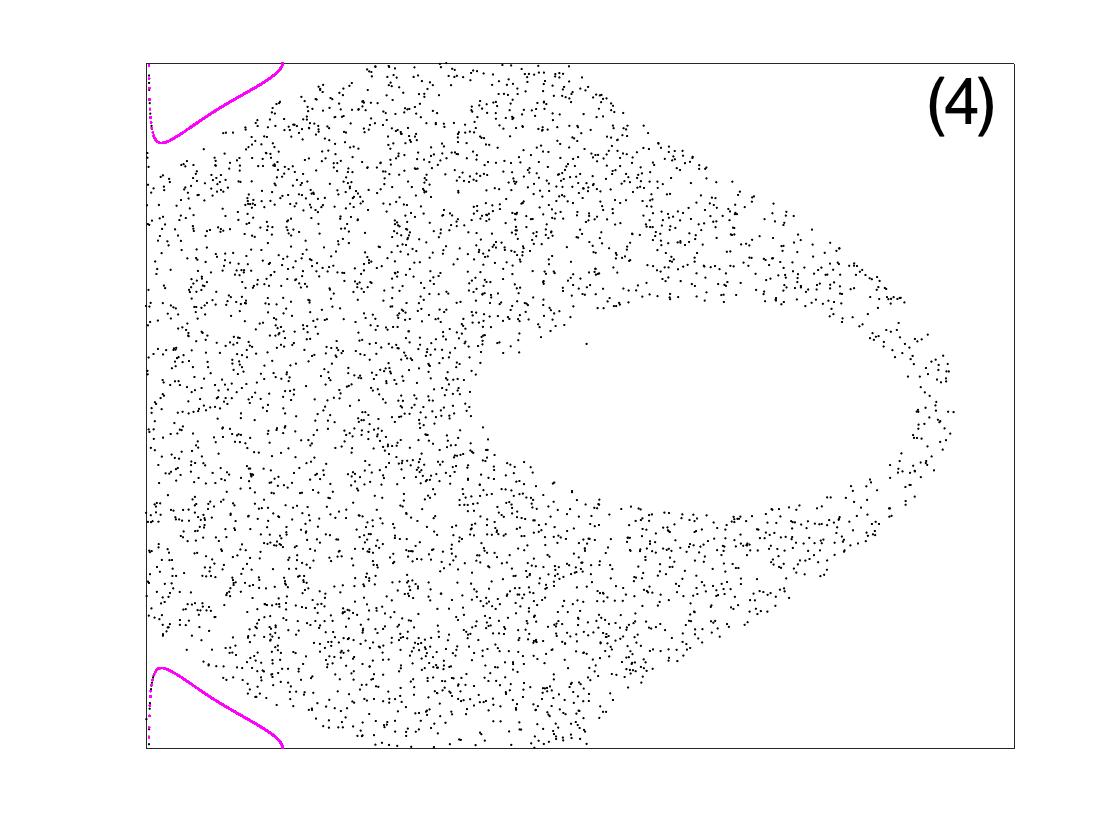}};
\node[right=0mm of f] (i)  {\includegraphics[width=.19\textwidth, trim=20mm 2mm 12mm 9mm, clip]{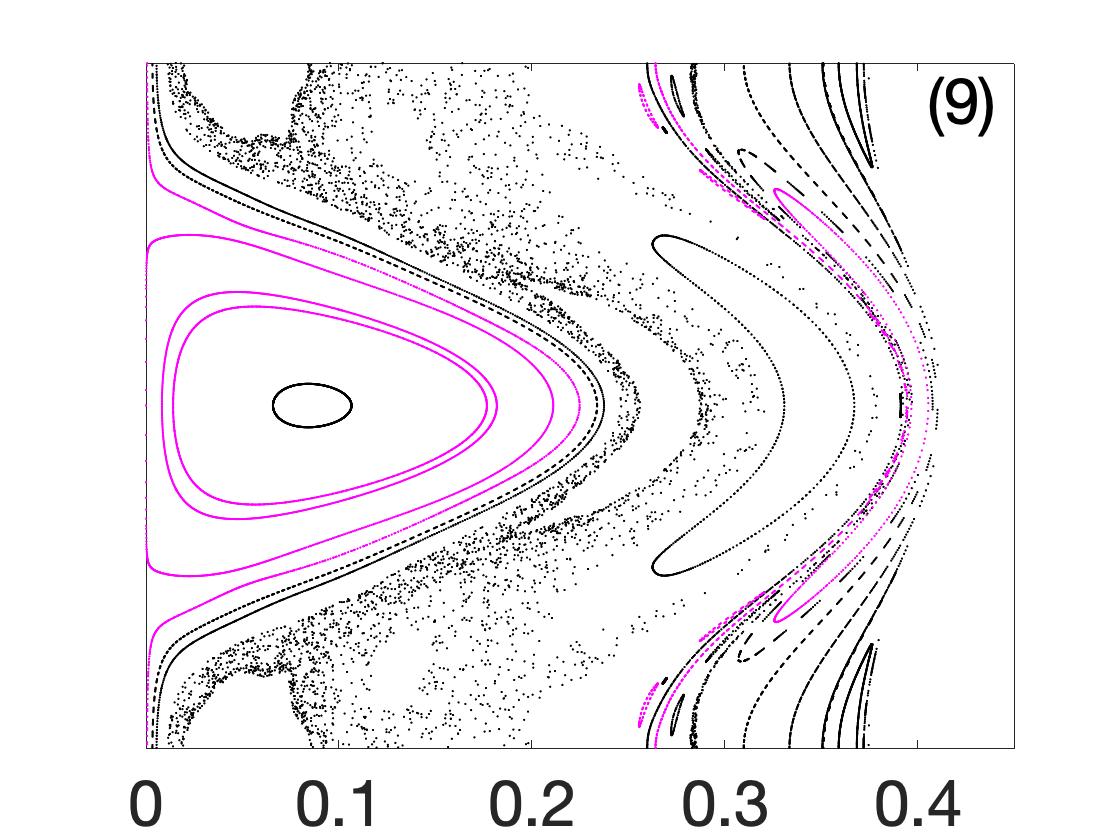}};
\node[above=0mm of i] (h)  {\includegraphics[width=.19\textwidth, trim=20mm 12mm 12mm 9mm, clip]{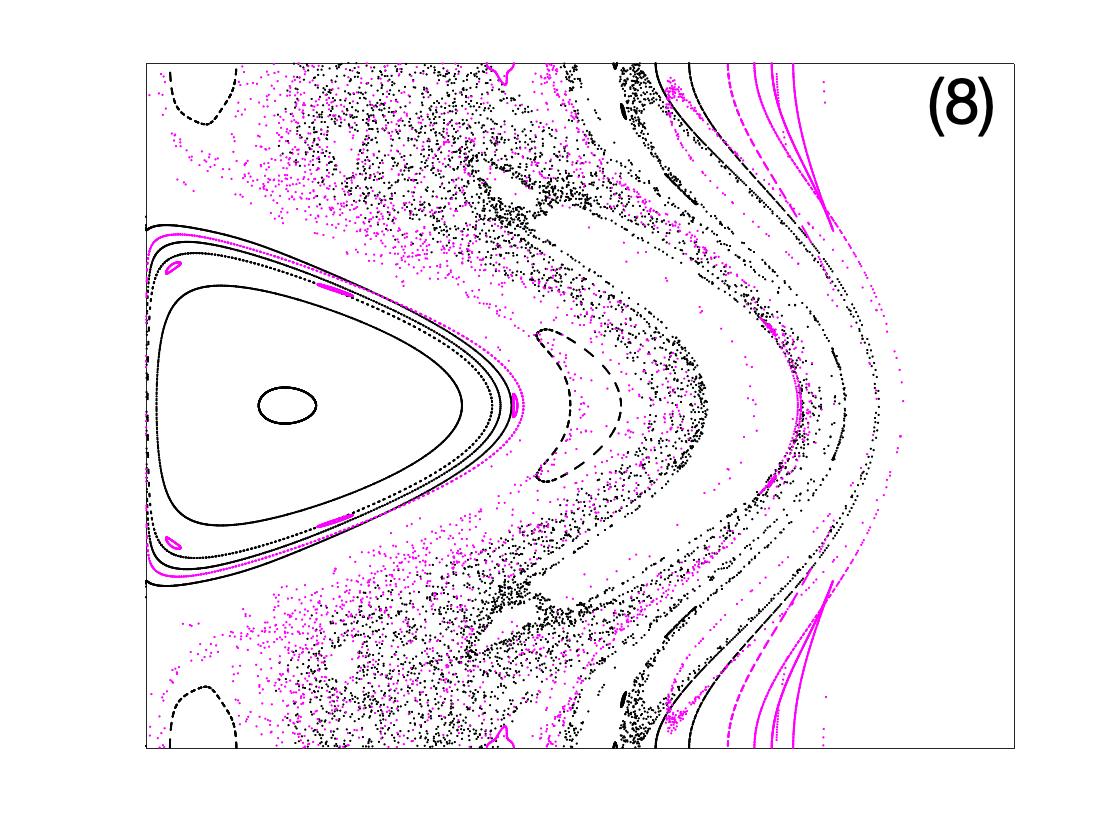}};
\node[above=0mm of h] (g)  {\includegraphics[width=.19\textwidth, trim=20mm 12mm 12mm 9mm, clip]{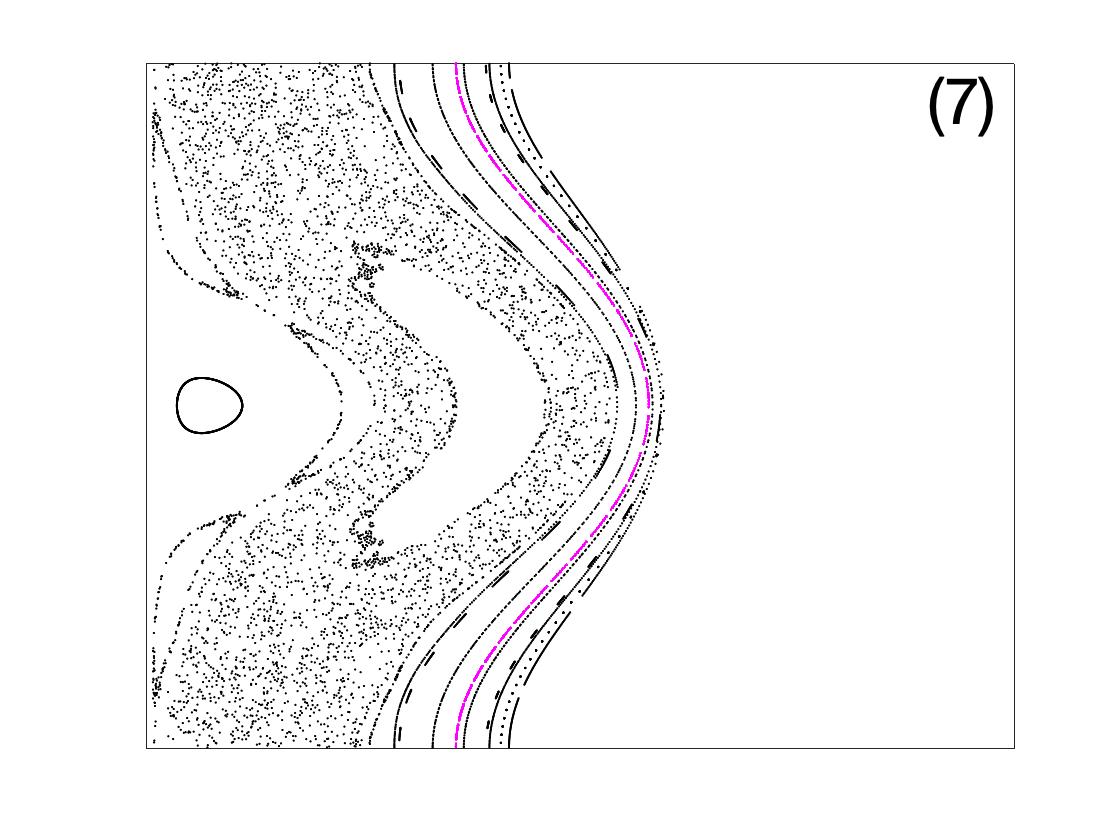}};
\node[above=0mm of b] (j)  {\includegraphics[width=.6\textwidth, trim=5mm 0mm 10mm 9mm, clip]{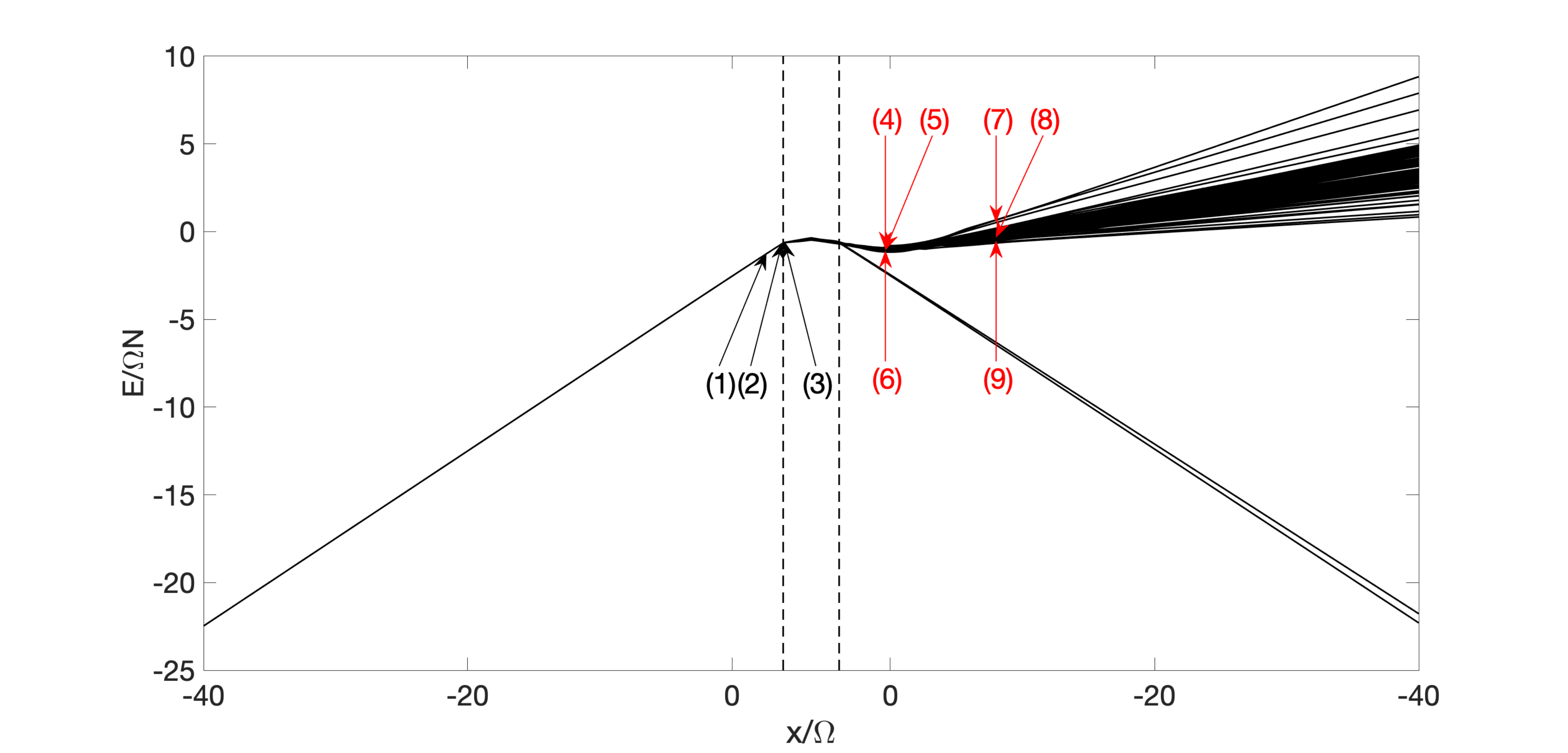}};
\end{tikzpicture}}
\caption{\label{f2}
Evolution of an ensemble of trajectories for (a) $u=-40$;  and (b) $u=-5$. 
The upper left plot in each set shows the time-dependent energies of the ensemble trajectories versus $x(t)$. The vertical dashed lines indicate the critical values $x_{S}$ and ${x_C}$ respectively. Arrows indicate $x$ values for which Poincar\'e sections are displayed for the forward sweep (black) and for the backward sweep (red).           
In the backward sweep the ensemble is no longer mono-energetic. In (a) just two $E$-sections are required because only two tori are involved, but in (b) a range of energy surfaces is occupied and three representative sections are shown.
Black points in the Poincar\'e sections are mono-energetic trajectories of the instantaneous Hamiltonian, while the magenta points show actual ensemble points within a narrow range of energies, evolving for several orbits under the instantaneous Hamiltonian.
}
\end{figure*}

\sect{Quasi-integrable scenario}  
Analyzing the Hamiltonian hysteresis scenario for $u=-40$, we find that the dynamics remains quasi-integrable throughout the process \cite{SMc}. This is illustrated in the Poincar\'e sections of Fig.~\ref{f2}(a) at representative $x$ values \cite{SMd}. All black points within a given section have the same energy~$E$. The magenta points indicate the quasi-static evolution, at each value of $x$, of those points in our actual ensemble with energy near $E$. Since energy is not a constant of motion, several $E$~sections are required in order to illustrate the ensemble at a given moment. The time-dependent energy for a some of the ensemble-trajectories is plotted as well.  
During the forward sweep the evolving trajectories remain restricted to the initially occupied torus $\bm{I}^{0}$. Beyond $x=x_S$ the occupied torus \textit{merges} with another (empty) torus $\bm{I}^{1}$. When $x=x_S$ is crossed again on the backward sweep, the ensemble splits between the two tori $\bm{I}^{0,1}$. Subsequently the two sub-ensembles evolve to different energies, as implied by Einstein-Landau adiabaticity, namely $E(t)=H[\bm{I}^{k};x(t)]$. Those trajectories that come back to the initial torus $\bm{I}^0$ contribute to $P(x_0)$. Their fraction can be determined by the Kruskal-Neishtadt-Henrard theorem \cite{Kruskal,Neishtadt1,Henrard,Neishtadt2,LC,dimer}.   

\sect{Passage through chaos scenario}  
Analyzing the Hamiltonian hysteresis scenario for $u=-5$, we observe in the Poincar\'e sections of Fig.~\ref{f2}(b) that for $x>x_C$ the initially regular orbits around an energy minimum merge into a chaotic zone \cite{SMc}. The ensemble trajectories spread ergodically throughout this chaotic zone, which is a section of a three-dimensional phase space region within the energy shell. On the backward sweep, the ergodized ensemble emerges from chaos step by step, gradually breaking into a range of many different integrable tori. Each final torus corresponds to a pair of adiabatically conserved action variables, which at late times can be identified as the occupation numbers $n_{1,3}$. 

The scatter of points in Fig.~\ref{f1}(b) is thus the direct experimental signal of the spreading of the ensemble into many tori after passage through chaos, while Fig.~\ref{f1}(a) shows the two final tori of the integrable scenario. The larger dispersion in phase space due to chaotic ergodization is responsible for the much lower return probabilities $P(x)$ for $x>x_C$ in Fig.~\ref{f1}(c).

\sect{Energy spreading}  
Had all tori formed at the same detuning $x=x_C$ in the reverse sweep, one would expect that the final distribution in Fig.~\ref{f1}(b) would trace a single contour in the $n_{1,3}$ plane and the energy would spread according to the formula $E(t)=H[\bm I;x(t)]$, where the actions belong to the surface $H[\bm I;x_C]=E_C$, with $E_C=H[\bm{I^0};x_C]$ that corresponds to the initially occupied torus. However, the transition from chaos to integrability is {\em gradual}, occurring over an intermediate mixed-phase-space interval during which integrable tori are formed one by one within the chaotic sea. Accordingly, each torus joins a different energy shell as it becomes Einstein-Landau-adiabatic. Some of these newly formed tori may later merge into chaos again, so that a trajectory may undergo several transitions from Ott to Einstein-Landau adiabaticity and back before it finally emerges from chaos. Thus the action distribution on the final exit from chaos is observed to have an {\em intrinsic} energy width due to the gradual nature of the transition, even if the sweep is infinitely slow.

After the final exit from chaos, energies spread out adiabatically according to the formula $E(t)=H[\bm{I};x(t)]$, where the actions $\bm{I}$ are conserved. This Einstein-Landau-adiabatic spreading of energy is \emph{not} irreversible. The energy contours shown in Fig.~\ref{f1}(b) are those of the Hamiltonian just after exit from chaos, however (see \cite{SMe}). Indeed, the plotted $n_{1,3}$ distribution in Fig.~\ref{f1}(b) corresponds to a broadened energy shell whose width decreases as the sweep becomes more adiabatic, but saturates to a finite value in the limit of an infinitely slow sweep (Fig.~\ref{f1}(d)). Between this intrinsic energy width and ergodic spreading along energy contours, the final ensemble after Hamiltonian hysteresis with chaos has the substantial entropy shown in Fig.~\ref{f1}(d) \cite{SMe}.

If the sweep is slow, but not strictly adiabatic, there will be additional energy spreading $\Delta E$. Under Einstein-Landau adiabaticity the spreading is negligible ($\sim \exp(-1/|\dot{x}|)$, but for fully developed chaos Ott's theory predicts diffusion in energy, namely $\Delta E^2 \propto \dot{x}^2 t^{\alpha}$ with $\alpha=1$, as opposed to the transient ballistic value $\alpha=2$. The duration of our `sweep through chaos' is $t\propto T=(|x_I|+x_0)/|\dot x|$, hence we expect $\Delta E \propto |\dot{x}|^a$  with $a = 1 - \alpha/2$. The numerical data of Fig.~\ref{f1}(d) fits the value of $a\sim 0.2$  which is between the $a=0$ of ballistic dynamics and the $a=0.5$ that would be expected in the diffusive case, before it levels off. The plateau that is observed for small $\dot x$ (large $T$) indicates that the quasi-static regime has been reached.

\sect{Conclusion and outlook} 
Experiments like those we have simulated will be able to show the decisive role of chaos right from the microscopic onset of irreversibility in small isolated systems, by observing the dramatic dependence of final $n_{j}$ distributions and the associated return probability $P$ on $x_0$, on $u$, and on sweep rate. Quantum corrections to the semiclassical results at smaller particle numbers warrant future investigation.

\acknowledgments{RB and JRA acknowledge support from State Research Center OPTIMAS and the Deutsche Forschungsgemeinschaft (DFG) through SFB/TR185 (OSCAR), Project No. 277625399. AV and DC acknowledge support by the Israel Science Foundation (Grant  No. 283/18).}

\clearpage

\onecolumngrid

\pagestyle{empty}
\renewcommand{\thefigure}{S\arabic{figure}}
\setcounter{figure}{0}

\Cn{
{\Large \bf Probabilistic Hysteresis in Integrable and Chaotic Isolated Hamiltonian Systems} \\
Ralf B\"urkle, Amichay Vardi, Doron Cohen and James R. Anglin \\
{SUPPLEMENTARY MATERIAL} \\ \
} 

This Supplementary Material provides technical details concerning the semi-classical dynamics of our Bose-Hubbard model, and also concerning the energy contours and entropies in Fig.~\ref{f1}. An additional figure similar to \ref{f1}(c) is also presented, to show that for many values of $u$ there exist both a separatrix-crossing threshold $x=x_S$ and a chaos-entry threshold $x=x_C > x_S$. In such cases the return probability $P(x)$ has two successive plateaus.

\section{The semi-classical Bose-Hubbard trimer}
\subsection{Phase space coordinates}
The classical Hamiltonian for a three-site Bose-Hubbard model is obtained from the quantum Hamiltonian operator (1) by replacing the operators $\hat a_j$, $\hat a_j^{\dagger}$ with complex variables
\begin{equation}
\begin{split}
\hat a_j &\rightarrow \alpha_j=\sqrt{n_j} e^{i \varphi_j}\\
\hat a_j^{\dagger} &\rightarrow \alpha_j^*=\sqrt{n_j} e^{-i \varphi_j},
\end{split}
\end{equation}
yielding
\begin{equation}
H=\frac{x}{2}(n_1-n_3)+\frac{U}{2} \sum_{j=1}^{3}n_j^2-\Omega \sum_{j=1}^{2}\sqrt{n_{j+1}n_j} \cos(\varphi_{j+1}-\varphi_j).
\end{equation}
If we then change to a slightly different set of canonical coordinates
\begin{equation}
\begin{split}
&q_1=\varphi_1-\varphi_3, \quad p_1=n_1\\
&q_2=\varphi_2-\varphi_3, \quad p_2=n_2\\
&q_3=\varphi_3, \quad p_3=\sum_{j=1}^{3}n_{j}
\end{split}
\end{equation}
then we can effectively eliminate one degree of freedom, because the total particle number $N=p_{3}$ is conserved and the common phase factor among all $\alpha_j$ is irrelevant. This yields the Hamiltonian in a four-dimensional phase space:
\begin{equation}
H=\frac{x(t)}{2}(2p_1+p_2-N)+\frac{U}{2}\left[(p_1+p_2)^2-N^2\right]-\Omega \left[\sqrt{p_1 p_2}\cos(q_2-q_1)+\sqrt{p_2(N-p_1-p_2)}\cos(q_2)\right]
\end{equation}
with the equations of motion
\begin{equation}
\begin{split}
\dot q_1&=-\frac{\Omega}{2}\left[\sqrt{\frac{p_2}{p_1}}\cos(q_2-q_1)-\sqrt{\frac{p_2}{N-p_1-p_2}} \cos(q_2)\right]+U(2p_1+p_2-N)+x\\
\dot p_1&=-\Omega \sqrt{p_1 p_2}\sin(q_1-q_2)\\
\dot q_2&=-\frac{\Omega}{2}\left[\sqrt{\frac{p_1}{p_1}}\cos(q_2-q_1)-\frac{N-p_1-2p_2}{\sqrt{p_2(N-p_1-p_2)}} \cos(q_2)\right]+U(p_1+2p_2-N)+\frac{x}{2}\\
\dot p_2&=\Omega\left[ \sqrt{p_1 p_2}\sin(q_1-q_2)-\sqrt{p_2(N-p_1-p_2)}\sin(q_2)\right]\;.
\label{eq:eqm}
\end{split}
\end{equation}

Evolving a single phase space point under (\ref{eq:eqm}) is classical evolution. The semi-classical approximation to the quantum mechanical evolution consists in evolving a positive Wigner function by flowing it along classical orbits, so that it evolves as a classical phase space probability distribution under the Liouville equation. Our results are based on classical dynamics in the sense that we evolve an initial ensemble under (\ref{eq:eqm}), but the conclusions we draw are all about the semi-classically approximated Wigner function which this ensemble samples, and hence represent a controlled approximation to quantum many-body dynamics which should be accurate for attainably large particle numbers $N$. 

In the integrable scenario ($u=-40$) the ensemble in the Poincar\'e sections is very close to the right edge of the sections. For better graphical representation we can introduce new canonical coordinates
\begin{align}
q&=\tan^{-1}\left(\frac{n_1}{\sqrt{N^2-n_1^2}} \cos(\varphi_1-\varphi_3)\right)\nonumber\\
p&=-\sqrt{N^2-n_1^2} \sin(\varphi_1-\varphi_3)\;.
\end{align}
that move the region of interest from the right edge to the center of each section. These coordinates will be used in the following section, and a variant of Fig.~\ref{f2}(a) using these coordinates can be found in Fig.~\ref{fS3}.

\subsection{Stationary points}
The ensembles that we study do not actually include any stationary points under (\ref{eq:eqm}), but when they are in the integrable regime they do orbit around certain stationary points. Since it is difficult to visualize motion in a four-dimensional phase space, identifying the instantaneous stable and unstable stationary points of (\ref{eq:eqm}) is useful in helping to understand the way our system behaves. In terms of the complex variables introduced above the equations of motion $i\dot \alpha_j=\partial H/ \partial \alpha_j^*$ read
\begin{equation}
i \begin{pmatrix} \dot \alpha_1\\ \dot \alpha_2 \\ \dot \alpha_3 \end{pmatrix}=\begin{pmatrix} U |\alpha_1|^2+\frac{x}{2}& -\frac{\Omega}{2}&0\\ -\frac{\Omega}{2}&U|\alpha_2|^2 & -\frac{\Omega}{2} \\ 0 & -\frac{\Omega}{2}& U |\alpha_3|^2-\frac{x}{2} \end{pmatrix} \begin{pmatrix} \alpha_1 \\ \alpha_2 \\ \alpha_3 \end{pmatrix}=H_0 \bm \alpha.
\end{equation}
To find the stationary points we set $i \dot{\bm \alpha}=\mu \bm \alpha$, where $\mu$ is the chemical potential. 

Since for stationary points the phase differences $q_1=\varphi_1-\varphi_3$, $q_2=\varphi_2-\varphi_3$ must be zero or $\pi$ (see Eq.~(\ref{eq:eqm})) and the overall phase is irrelevant, we can assume the $\alpha_j$ to be real. It then follows
\begin{equation}
\begin{split}
U \alpha_1^2+\frac{x}{2} - \frac{\Omega}{2}\frac{\alpha_2}{\alpha_1}&=\mu \\
U \alpha_2^2-\frac{\Omega}{2}\frac{\alpha_1}{\alpha_2}-\frac{\Omega}{2}\frac{\alpha_3}{\alpha_2}&=\mu \\
U \alpha_3^2-\frac{x}{2} - \frac{\Omega}{2}\frac{\alpha_2}{\alpha_3}&=\mu
\end{split}
\end{equation}
Solving these equations together with $\alpha_1^2+\alpha_2^2+\alpha_3^2=N$ gives the stationary points as well as their energies in dependence of $x$, which we show in Fig.~\ref{fig:IntScen} for $u=-40$ and Fig.~\ref{fig:ChaosScen} for $u=-5$.

The stability of the stationary points is then determined by canonically linearizing the grand canonical Hamiltonian $H-\mu N$ (since $\alpha \rightarrow \alpha + \delta \alpha$ changes the total particle number) around the stationary points $\alpha_0$. Diagonalizing the linearized Hamiltonian yields the Bogoliubov-de Gennes eigenfrquencies $\omega_B$, which are in general complex. The existence of eigenfrequencies with non-zero imaginary parts signals a \textit{dynamically unstable} stationary point. The existence of negative real eigenfrequencies for modes with positive Bogoliubov-de Gennes norm indicates \textit{energetic instability}.

\subsection{Ensemble orbits around stationary points}
\begin{figure}
\subfloat[]{\includegraphics[width=0.3\textwidth]{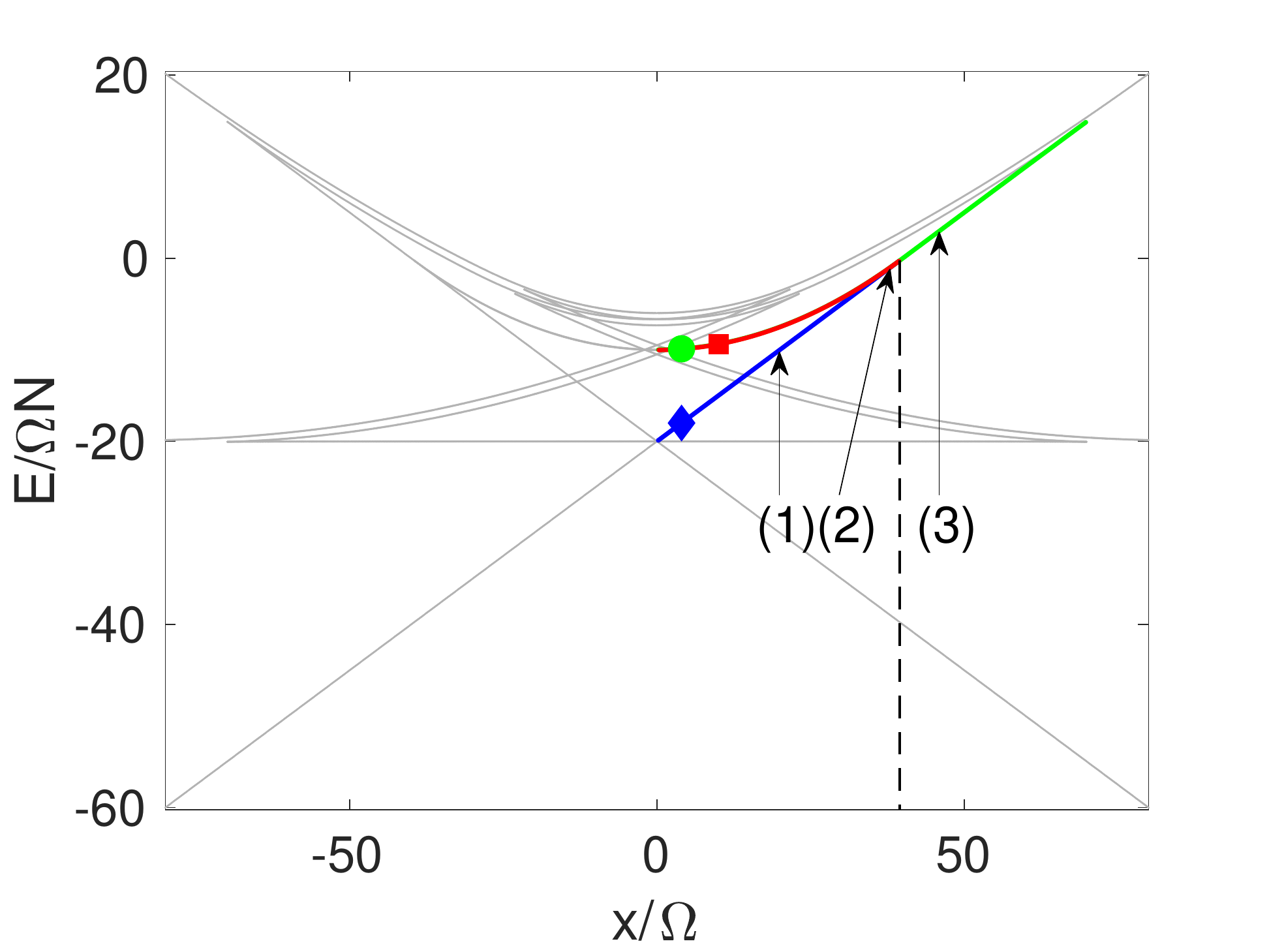}}
\subfloat[]{\includegraphics[width=0.225\textwidth]{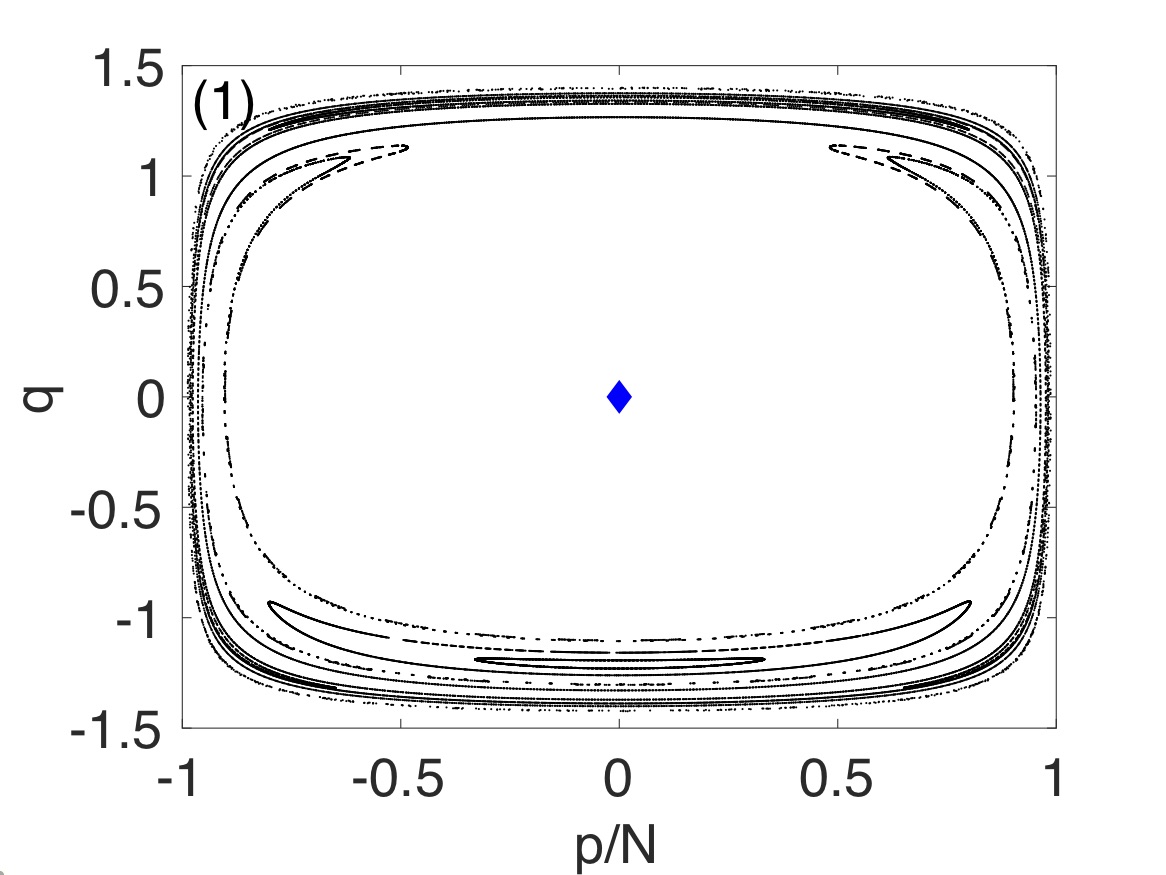}}
\subfloat[]{\includegraphics[width=0.225\textwidth]{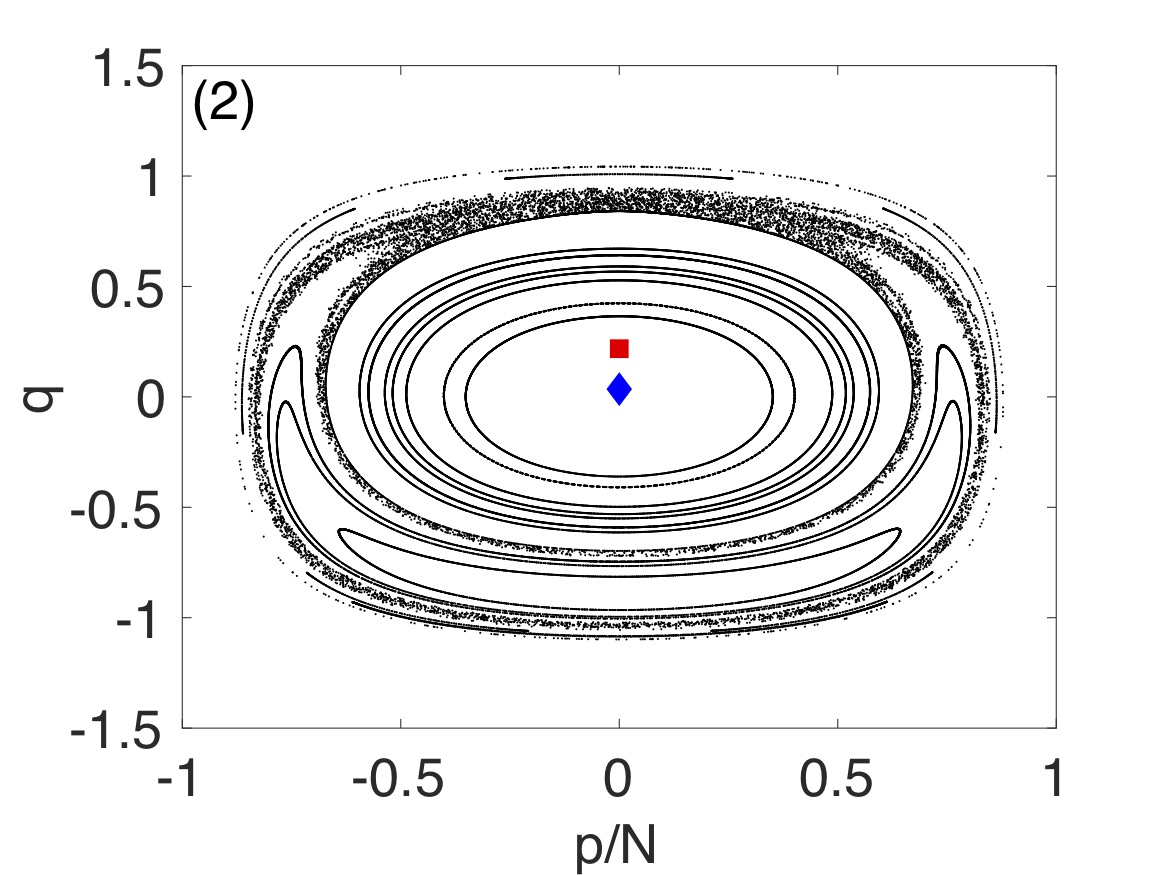}}
\subfloat[]{\includegraphics[width=0.225\textwidth]{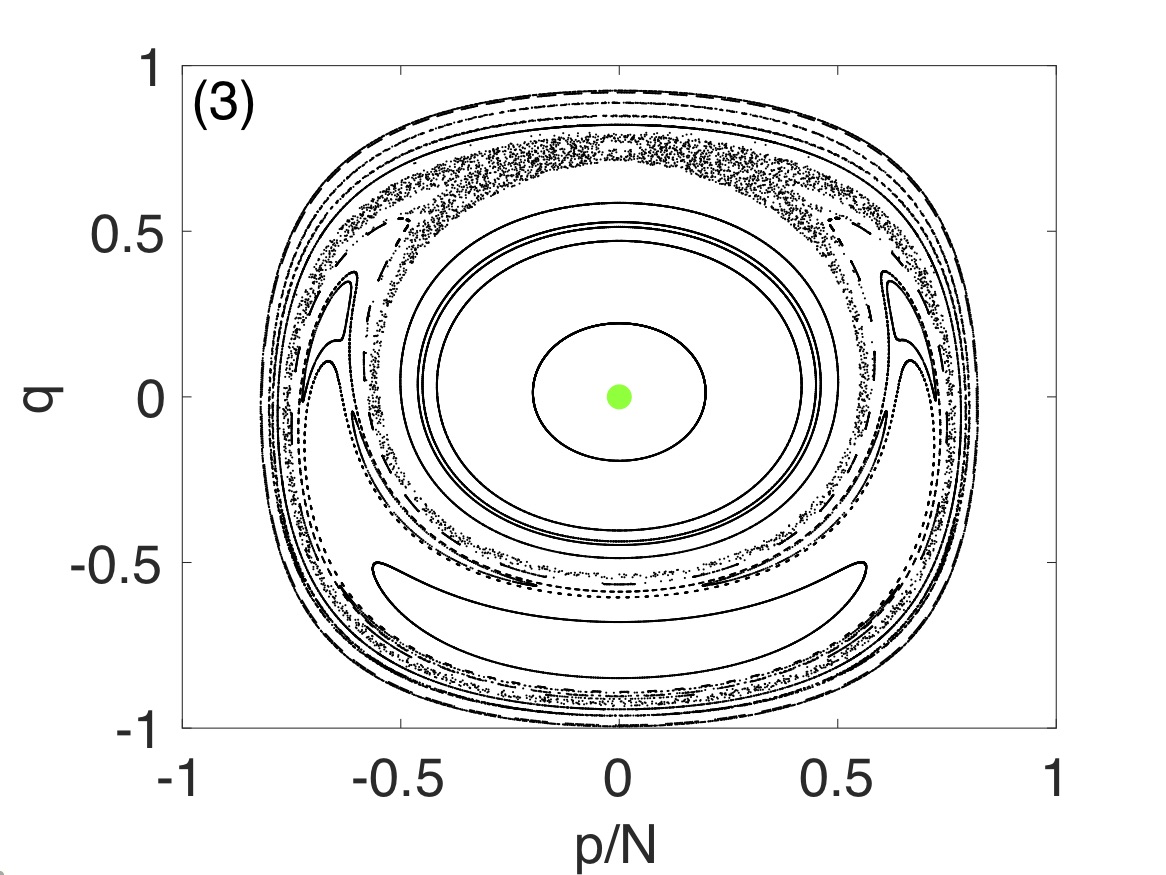}}
\caption{The energy of the stationary points is plotted as a function of the detuning in the left panel (grey lines) for $u=-40$. Colored lines and markers indicate the stationary points shown in panels (b)-(d). Poincar\'e sections of the quasi-static (fixed $x$) dynamics are taken at: (1)  $x<x_S$, (2) $x \lesssim x_S$ and (3) $x>x_S$. The followed energy minimum (blue line in (a)), marked by a blue diamond in (b) and (c) merges with a dynamically unstable saddle (red line in (a)), marked by a red square in (c), at $x=x_S$ (dashed line) and both disappear (c). In the following evolution the system follows the dynamically stable saddle (green line in (a)), marked by a green circle (d). In all cases except (d) the planes of section are $\varphi_{2}=\varphi_{3}$ and $E$ the energy of the included stationary point. In (d) the stationary point has $\varphi_{2}=\varphi_{3}+\pi$, so this defines the Poincar\'e section.}
\label{fig:IntScen}
\end{figure}

In the integrable hysteresis scenario for $u=-40$, the system initially orbits around an energy minimum, until at $x=x_S$ this energy minimum merges with a dynamically unstable stationary point and both disappear, see Fig.~\ref{fig:IntScen}. This procedure is analogous to the vanishing of the separatrix in the dimer, when the stable and unstable fixed point merge, as discussed in \cite{dimer}. The ensemble then continues in regular motion, however, by orbiting a nearby dynamically stable energetic saddle stationary point (note that an energetic saddle in a higher dimensional phase space can still be dynamically stable, unlike in the more familiar two-dimensional case) indicated by a green line/green dot in Fig.~\ref{fig:IntScen}. Inspecting the Poincar\'e sections in Fig.~\ref{f2}(a), we see that the mechanism of hysteresis is thus similar to the one in the double well example and in the Bose-Hubbard dimer \cite{dimer}. While the adiabatic dynamics remains restricted to phase space tori, at the critical value of $x=x_S$ during the forward sweep two instantaneous tori \textit{merge}. When $x=x_S$ again on the backward sweep, the merged torus bifurcates again into two tori with different energies, orbiting different stationary states. After the merger, orbits that were initially confined to one of the merging tori quasi-ergodically explore the combined torus. Thus when the combined torus separates again in the reverse sweep of $x$, only a fraction $P(x_0)$ of the ensemble returns to the initial torus. The other orbits end up in the other daughter torus instead. Since each torus is represented by a point in the actions plane, the outcome of this separation is the observed binary distribution of Fig.~\ref{f1}(a).

\begin{figure}
\subfloat[]{\includegraphics[width=0.3\textwidth]{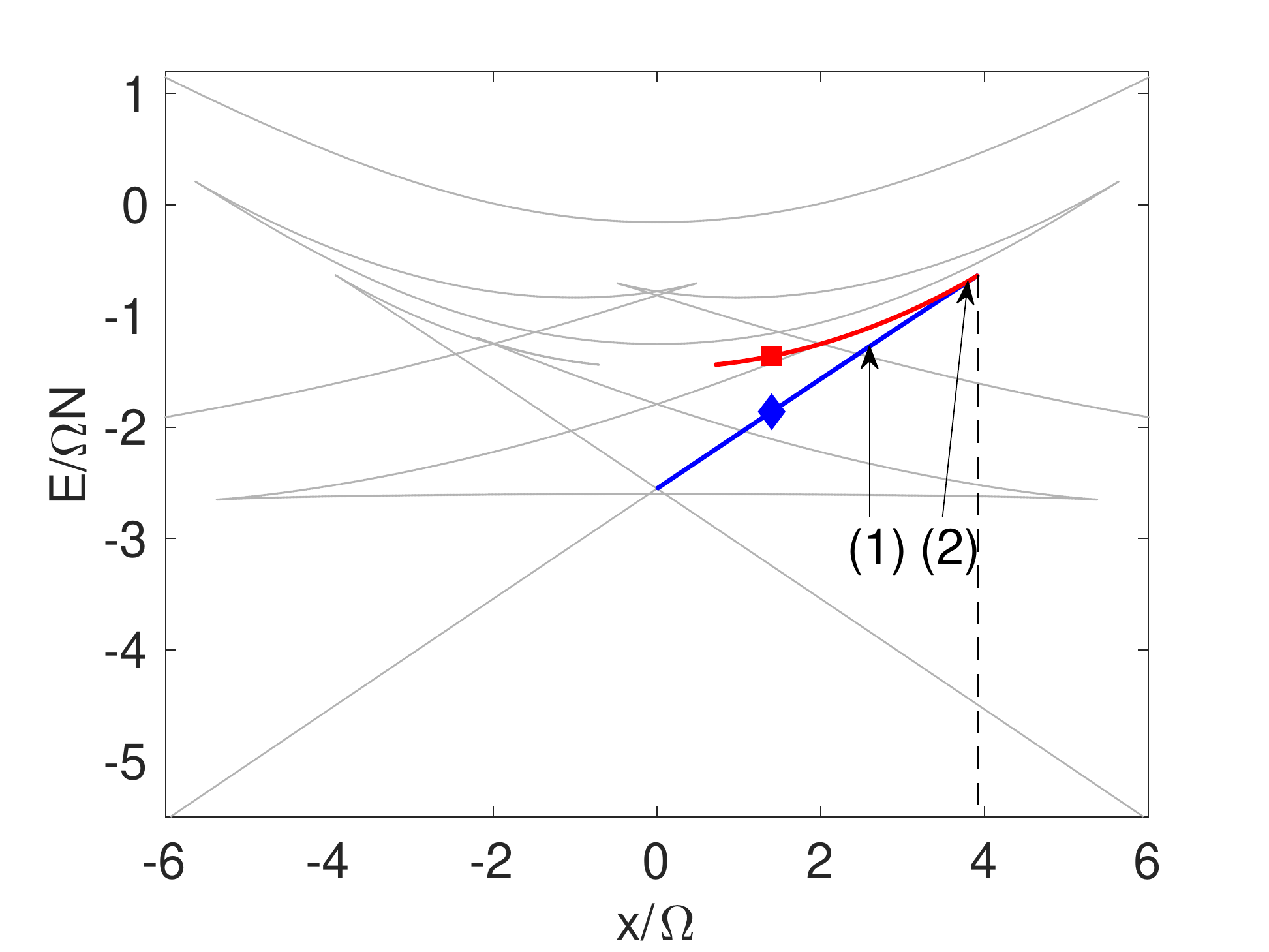}}
\subfloat[]{\includegraphics[width=0.225\textwidth]{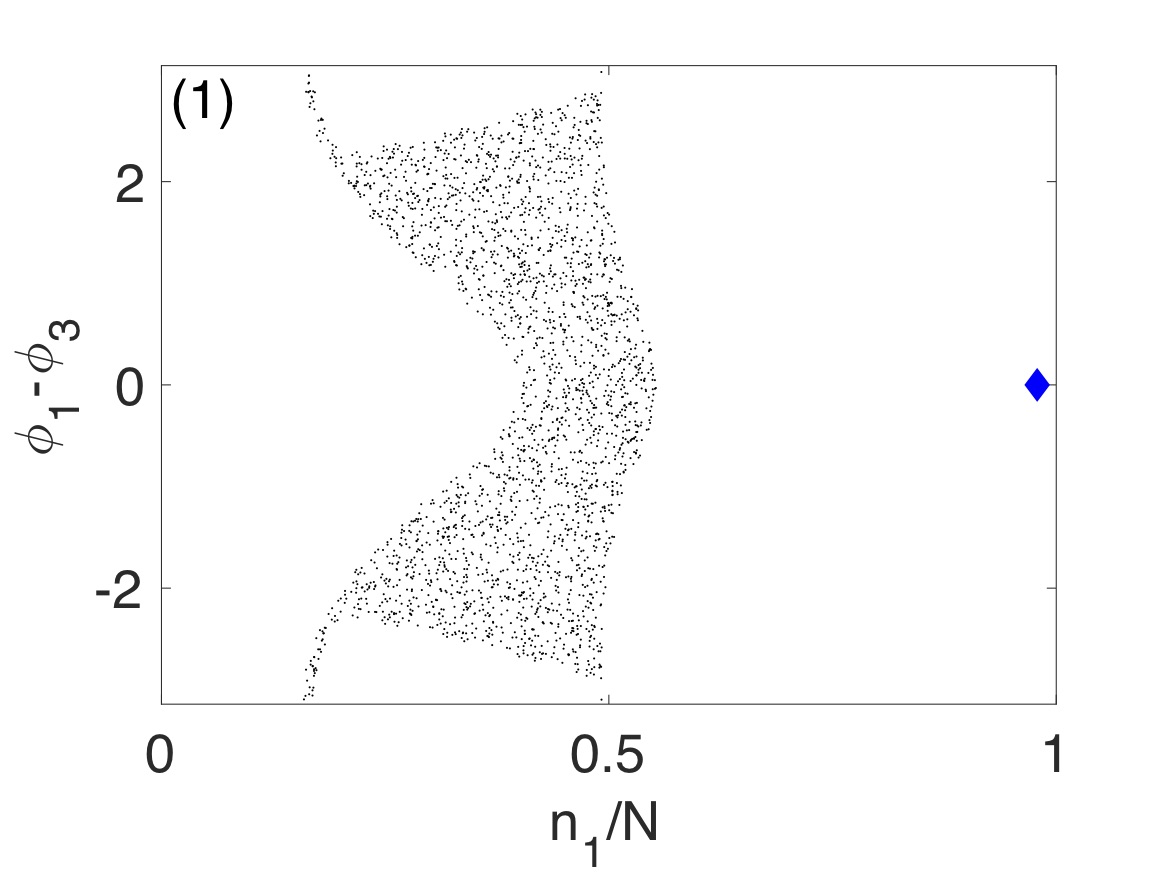}}
\subfloat[]{\includegraphics[width=0.225\textwidth]{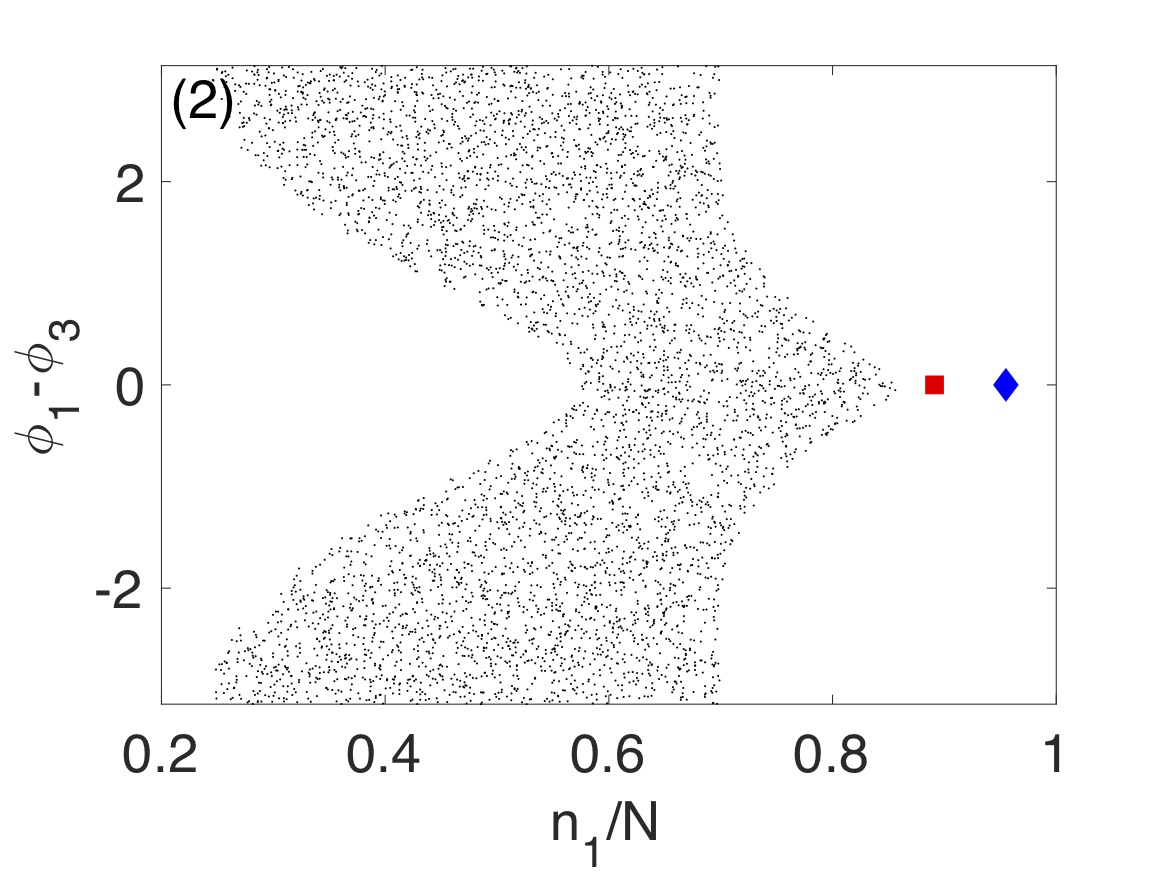}}
\caption{The energy of the stationary points is plotted as a function of the detuning in the left panel (grey lines) for $u=-5$. Colored lines and markers indicate the stationary points shown in panels (b) and (c). Poincar\'e sections of the quasi-static (fixed $x$) dynamics are taken at: (1)  $x<x_S$ and (2) $x \lesssim x_S$. The followed energy minimum (blue line in (a)), marked by a blue diamond in (b), merges with a dynamically unstable saddle (red line in (a)), marked by a red square in (c), at $x=x_S$ (dashed line). Since the saddle point is on the edge of the chaotic sea and there is no dynamically stable stationary point which the system could follow, the motion becomes chaotic (\textit{i.e.} $x=x_{C}$ is reached soon after $x=x_{S}$). The planes of section are $q_{2}=0$ and $E$ given by the energy of the included stationary point.}
\label{fig:ChaosScen}
\end{figure}

\begin{figure*}
\begin{tikzpicture}
\node (a) at (0,0) {\includegraphics[width=.19\textwidth, trim=0mm 0mm 18mm 9mm, clip]{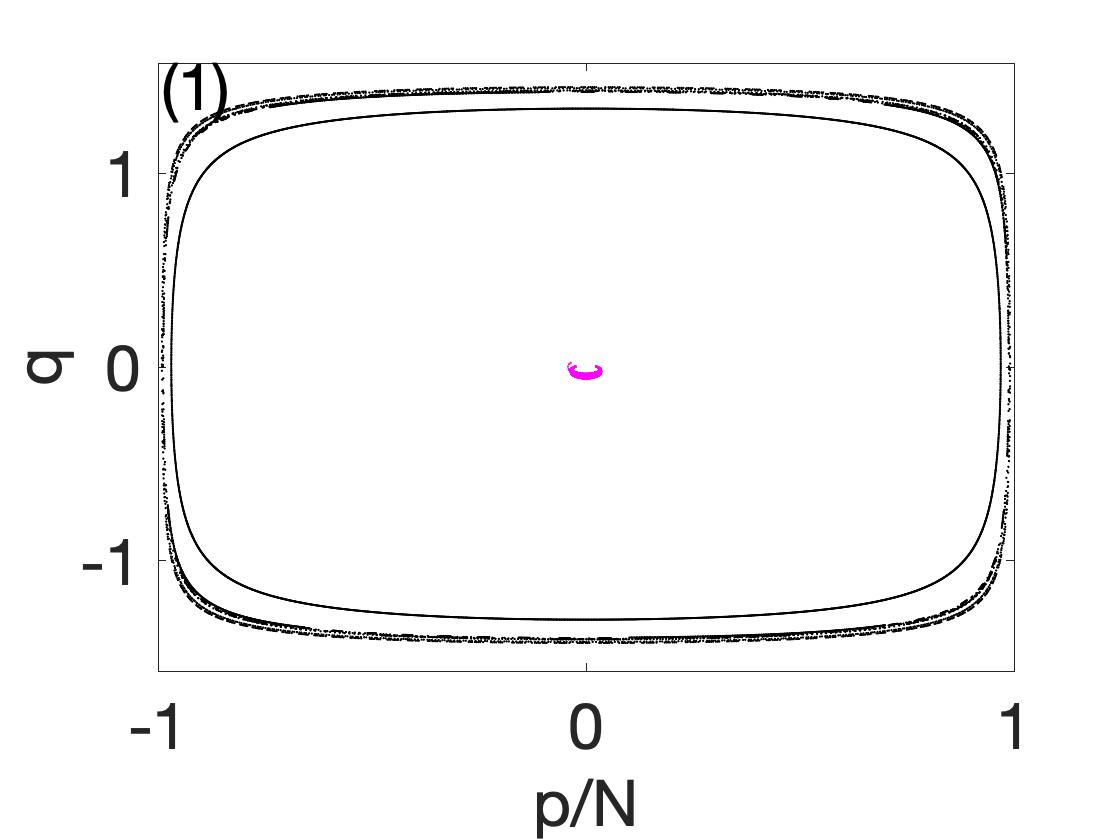}}; 
\node[ right=0mm of a] (b)  {\includegraphics[width=.19\textwidth, trim=25mm 12mm 18mm 9mm, clip]{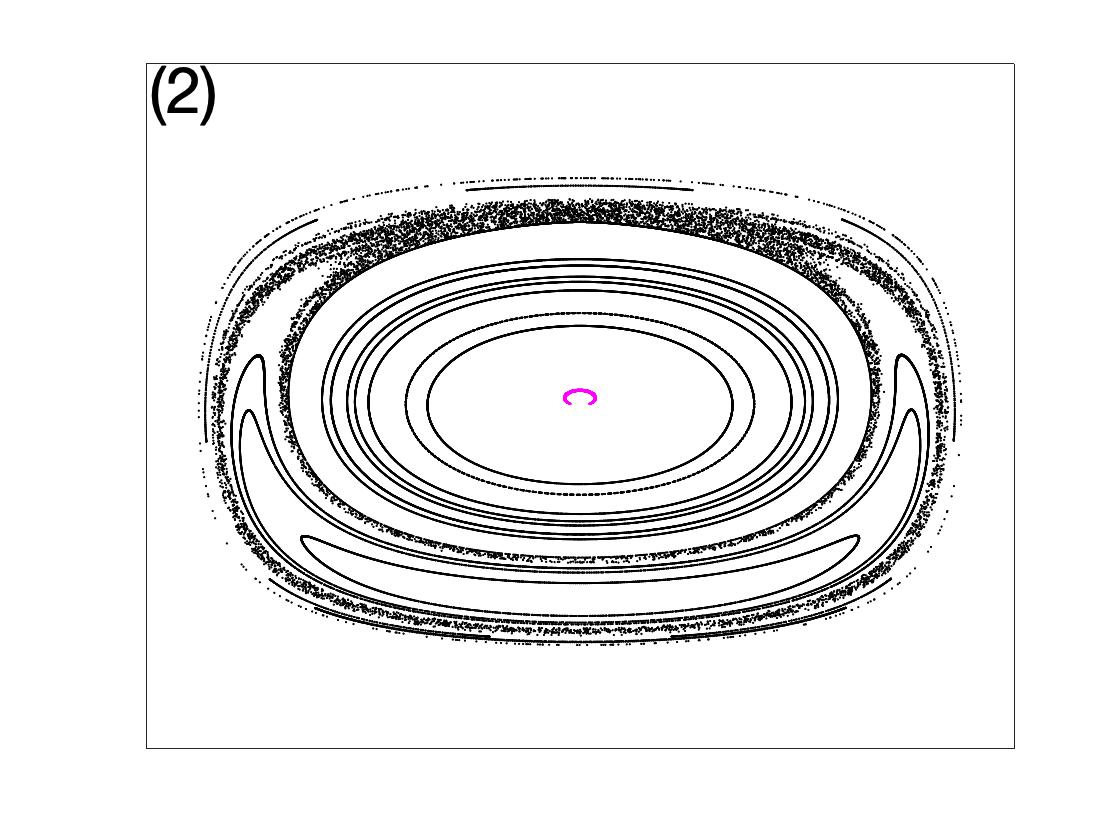}};
\node[right=0mm of b] (c)  {\includegraphics[width=.19\textwidth, trim=25mm 12mm 18mm 9mm, clip]{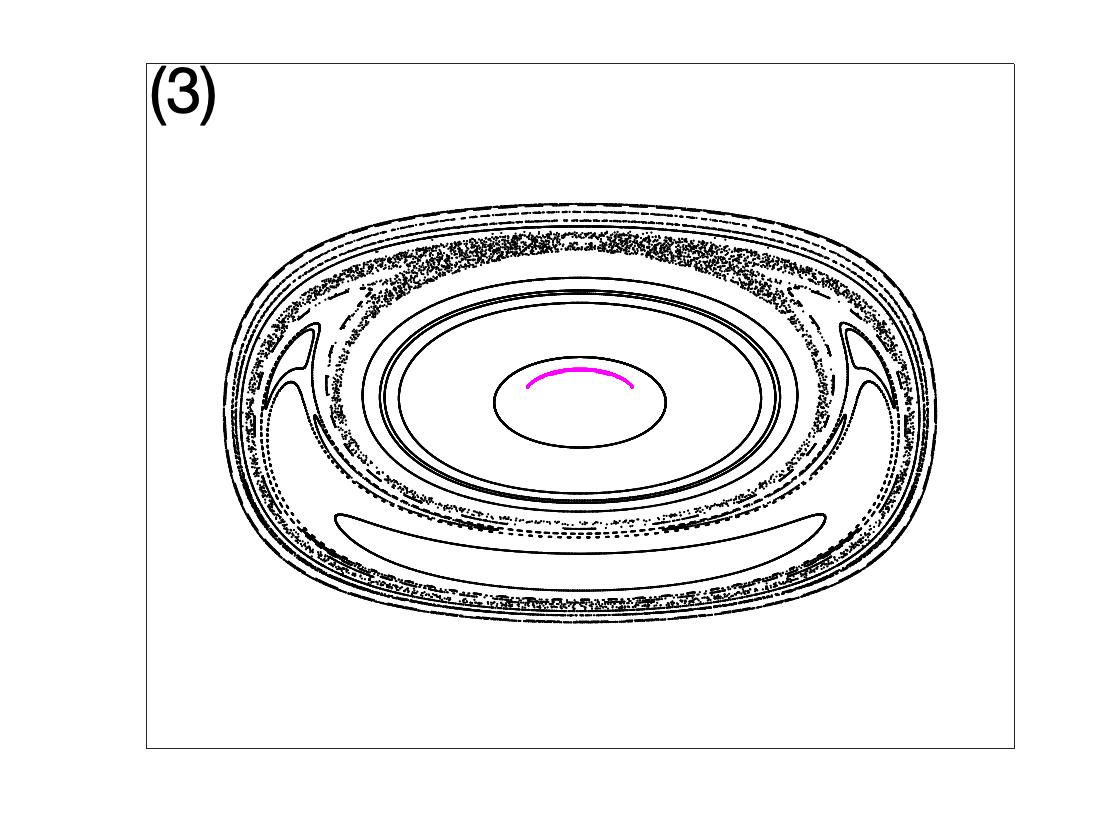}};
\node[right=0mm of c] (e)  {\includegraphics[width=.19\textwidth, trim=25mm 12mm 18mm 9mm, clip]{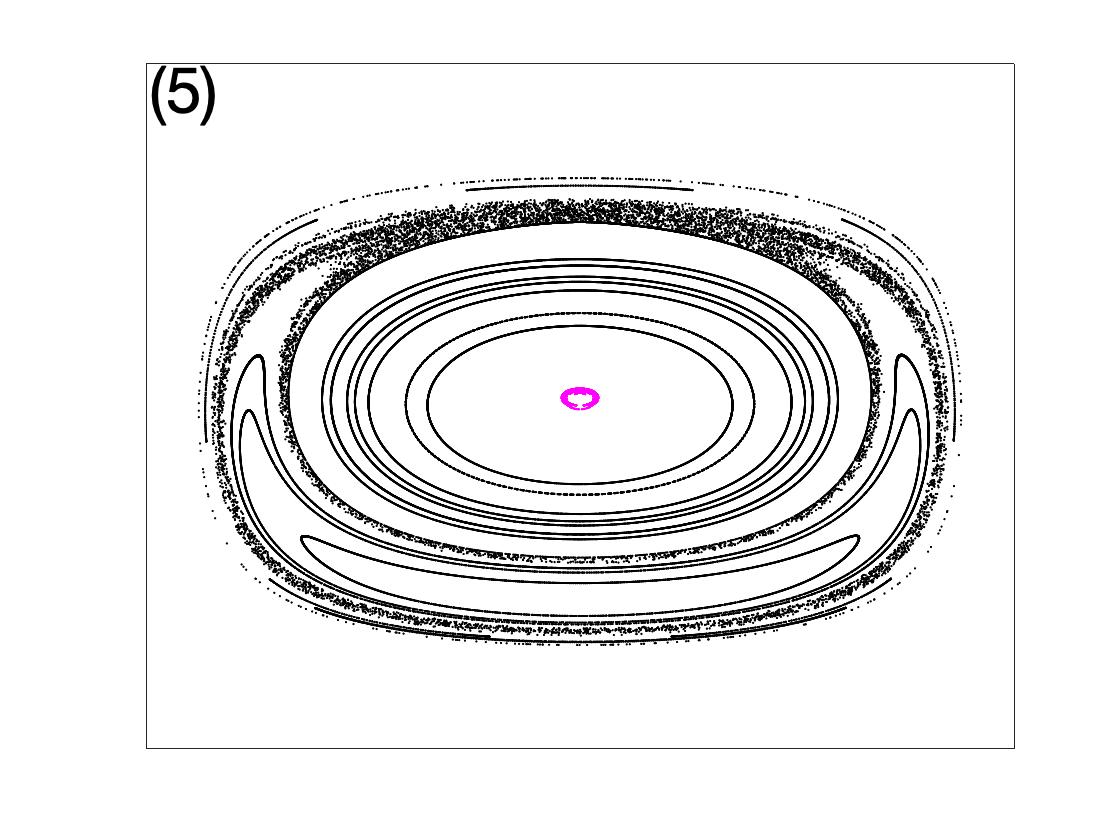}};
\node[right=0mm of e] (g)  {\includegraphics[width=.19\textwidth, trim=25mm 12mm 18mm 9mm, clip]{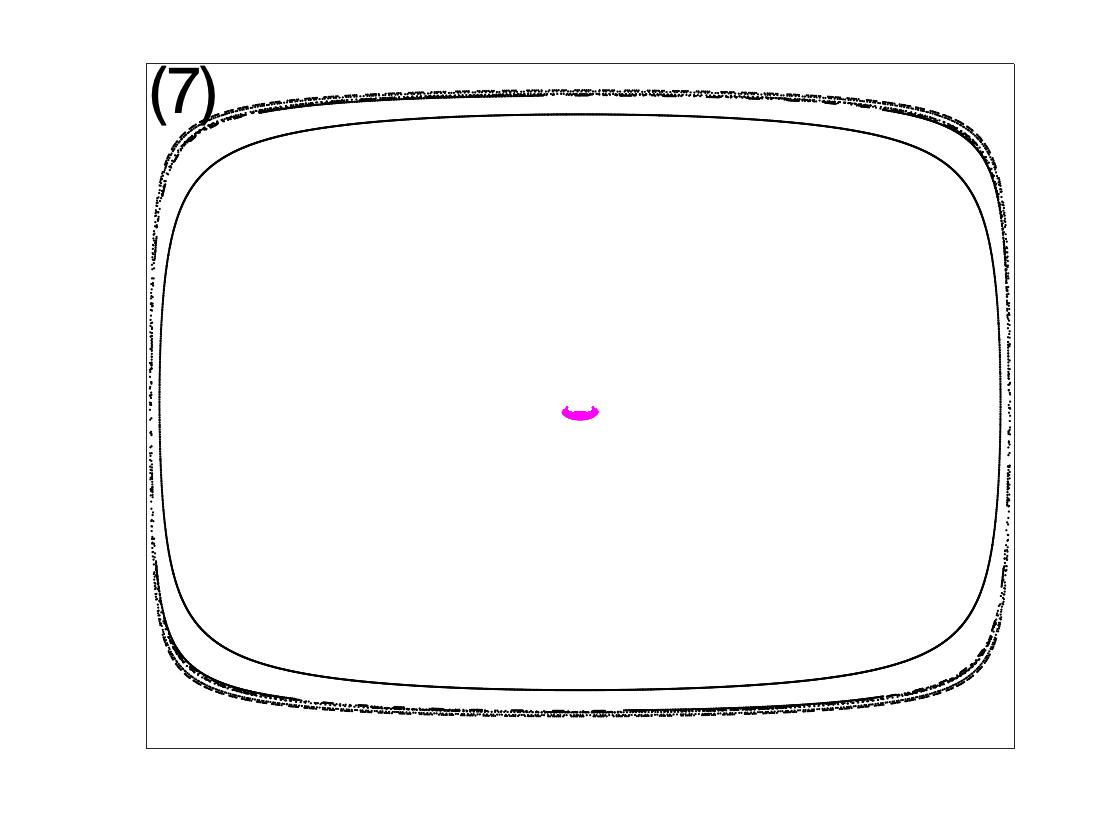}};
\node[above=0mm of e] (d) {\includegraphics[width=.19\textwidth, trim=25mm 12mm 18mm 9mm, clip]{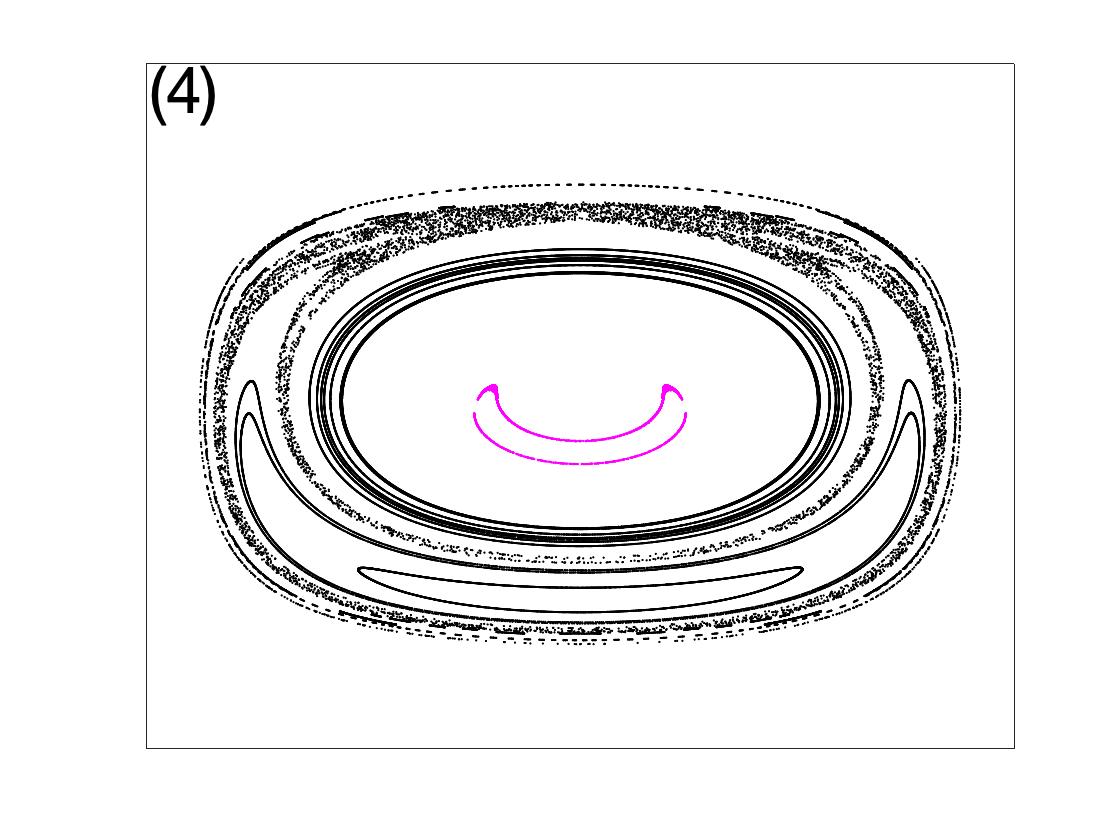}};
\node[right=0mm of d] (f)  {\includegraphics[width=.19\textwidth, trim=25mm 12mm 18mm 9mm, clip]{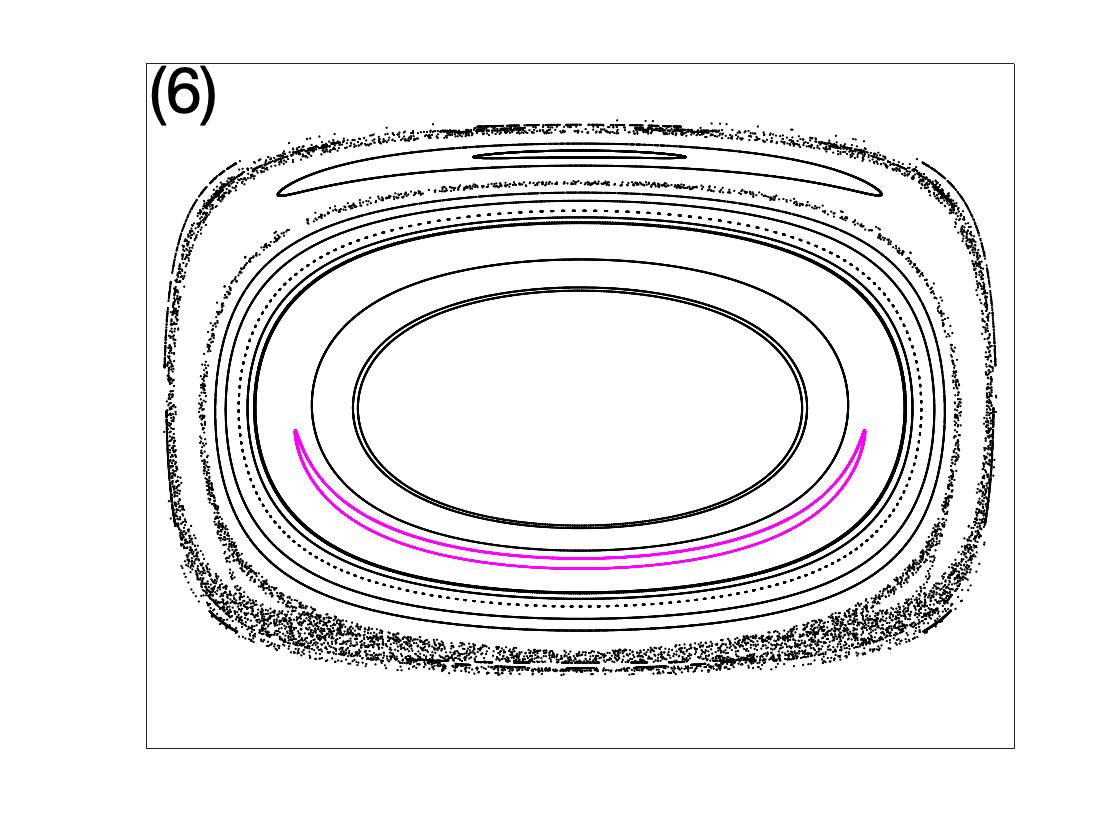}};
\node[above=0mm of b] (h)  {\includegraphics[width=.6\textwidth, trim=5mm 0mm 10mm 7mm, clip]{Fig2a.pdf}};
\end{tikzpicture}
\caption{\label{fS3}
Same as Fig.~\ref{f2}(a), but in the new coordinates $(q,p)$ in which the region of interest is in the middle of the section.
}
\end{figure*}

In the passage through chaos scenario with $u=-5$, it is again clear from Fig.~\ref{f2}(b) and Fig.~\ref{fig:ChaosScen} that for $x<x_C$ the ensemble orbits an energy minimum. At $x=x_C$ the energy minimum and a dynamically unstable stationary point merge and vanish, see Fig.~\ref{fig:ChaosScen}. In contrast to the $u=-40$ case, however, there is now no persistent integrable island near enough to the ensemble for it to continue in regular motion. The energetic barrier between the ensemble and the chaotic sea simply decreases until the ensemble enters chaos at $x=x_{C}$. After being ergodically dispersed throughout the chaotic sea, and being dispersed over a range of energies, the ensemble emerges from chaos on the return sweep into a wide range of different integrable tori.

\section{Poincar\'e sections}
The Poincar\'e sections shown in our Fig.~\ref{f2} illustrate the evolution of our ensemble (in magenta) around a particular time, as well as the range of possible evolutions under the instantaneous Hamiltonian at that time (in black). This presentation is intended to be analogous to the common procedure in two-dimensional phase spaces, of showing both system points and instantaneous energy contours. Since our phase space is four-dimensional, we show Poincar\'e sections. 

The Poincar\'e sections in each panel of Fig.~\ref{f2} are then constructed as follows. All points in a given section correspond to trajectories with the same energy, as marked in the upper left panels. A trajectory is sampled each time $q_{2}=0$ (sections 2,3,4 and 5 of Fig.~\ref{f2}(a), and sections 1,2, and 3 of Fig.~\ref{f2}(b)) or $q_{2}=\pi$ (all other sections). The alternatives $q_{2}=0,\pi$ are chosen in each case to ensure that the section does intersect our ensemble trajectories.

Note that the Poincar\'e sections shown in Fig.~\ref{f2} do not always cover the entire planes, but instead show only the portions that are near our ensemble for each $x$. Finally, our ensemble does not remain monoenergetic, but Poincar\'e sections show successive points in a trajectory of constant energy. Wherever Fig.~\ref{f2} has a stack of vertical panels, each panel includes magenta (ensemble) points from a narrow range of energies, and black points from a single energy in the middle of that range. In the integrable scenario there are only two narrow ranges of energies, so they are both shown in Fig.~\ref{f2}(a), but the three vertically stacked panels in Fig.~\ref{f2}(b) are representative selections from a larger energy range.

\section{Energy and entropy in the action plane}
\subsection{Energy}
Fig.~\ref{f1}(b) shows contours in the $(n_{1},n_3)$ plane of energy at the time $t_{ex}=4.26 \times 10^4 \hbar/\Omega$ ($x=9.96$), shortly after the ensemble with $\Omega T/\hbar=122667$ (green dots) has exited chaos in the return sweep from $x_0$ back to $x_I$. Since the $(n_1,n_3)$ points represent measurements at the time $t=+T$, however, when the Hamiltonian is quite different from what it was at $t_{ex}$, the meaning of these important energy contours needs explaining.

Our ensemble's dynamics is integrable at $t_{ex}$ and afterwards. Canonical action-angle variables $(I_a,\theta_a)$ and $(I_b,\theta_b)$ can therefore in principle be constructed. Constructing these variables as explicit functions of the original variables is in practice difficult, but because of Landau adiabaticity the actual values of the actions $(I_a,I_b)$ at $t_{ex}$ and at $t=T$ are the same. At $t=T$ the detuning $x$ is so large that the two occupation numbers $n_{1,3}$ are essentially identical with the action variables, simply because the coupling term in $H$ becomes off-resonant. Even without knowing the explicit functional form of the action variables at $t_{ex}$, therefore, we have the values of these variables at $t_{ex}$ numerically for every trajectory. Our large ensemble of trajectories means that we have a large sample of $(I_a,I_b)$ values at $t_{ex}$; as Fig.~1(b) shows, the sample is quite dense in the $(I_a,I_b)$ plane within our region of interest. Moreover we also know the energy of each of our trajectories for all times between $-T$ and $T$, and so in particular we know the energy at $t_{ex}$. Our large ensemble of trajectories therefore provides us a dense sampling of the energy at $t_{ex}$ as a function of $(I_a,I_b)$ at $t_{ex}$. We simply fitted a polynomial function in $I_a$ and $I_b$ to this sample (a simple quadratic function resulted in an excellent fit) and the energy contours in Fig.~1(b) are contours of this fitted function.
 
\subsection{Entropy}
Fig.~\ref{f1}(d) shows entropies associated with our final $n_{1,3}(T)$ distributions in the chaotic scenario with $u=-5$, for a wide range of different sweep rates $1/T$. These entropies were computed in the following way.

It is a numerically empirical fact that all our $n_{1,3}(T)$ distributions appear to be quite ergodic: points seem to be distributed randomly and evenly along the full extent of any energy contour at $t_{ex}$. What varies depending on $T$ is the width of energies over which the points are spread. For each final ensemble whose entropy is shown in \ref{f1}(d), therefore, we construct a \textit{kernel density estimate}: we fit the distribution of energies $E(t_{ex})$ to a smooth probability distribution $P_{fit}(E)$ made of superposed Gaussians, all Gaussians having the same width $b$, with their heights and positions as fitting parameters. This fitted distribution of energies was then used to define an ergodic probability distribution in four-dimensional phase space. The Boltzmann entropy of this distribution was then shown in Fig.~\ref{f1}(d).

Since our semiclassical evolution is Liouvillian, the fine-grained entropy of our ensemble cannot and does not increase over time. The entropy we compute is the coarse-grained entropy which is relevant to measurements of $n_{1,3}(T)$, which are the invariant action variables at late times, and which are also directly measurable in cold atom experiments.

\begin{figure}
\subfloat[]{\includegraphics[width=0.45\textwidth]{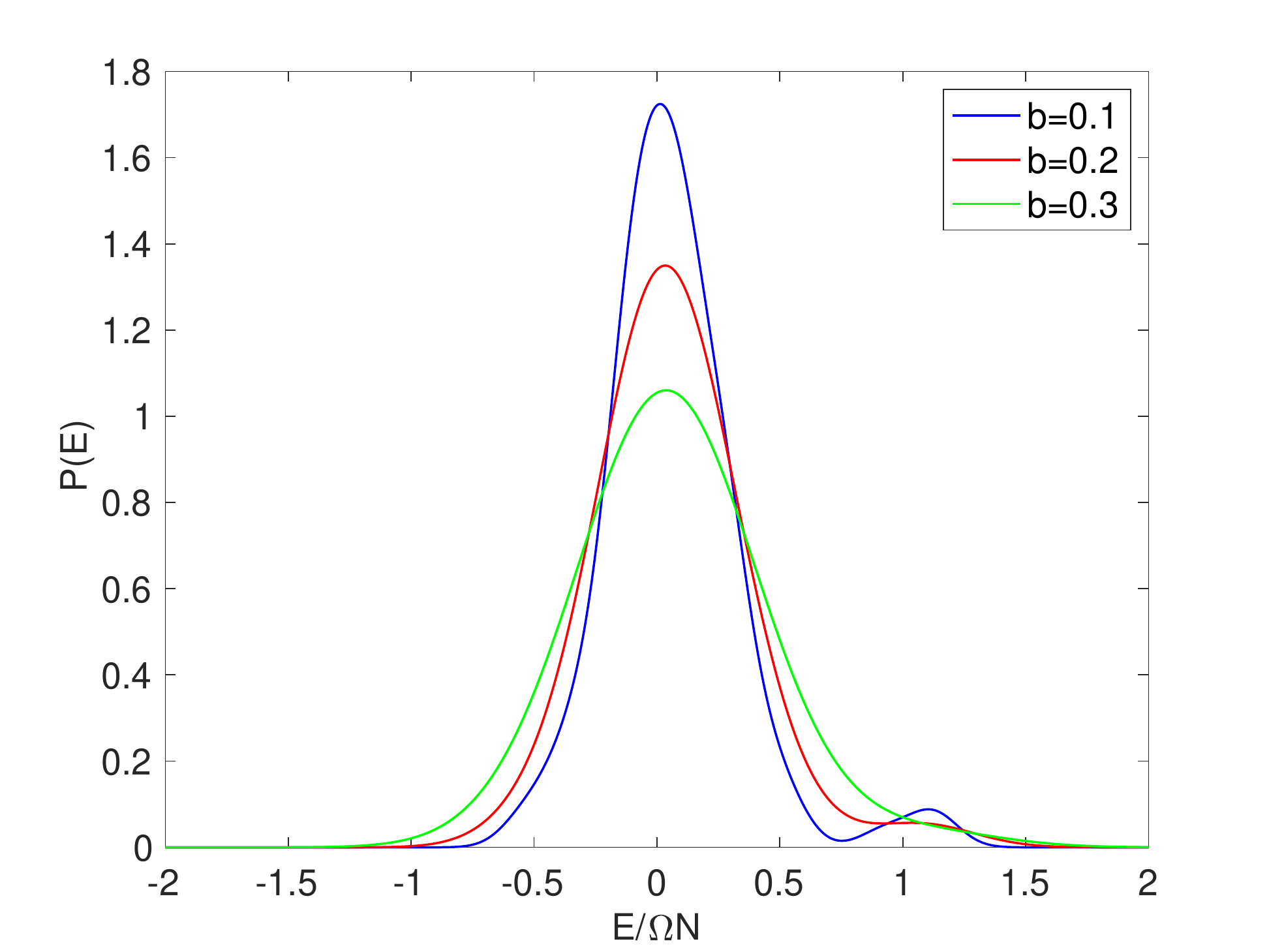}}
\subfloat[]{\includegraphics[width=0.45\textwidth]{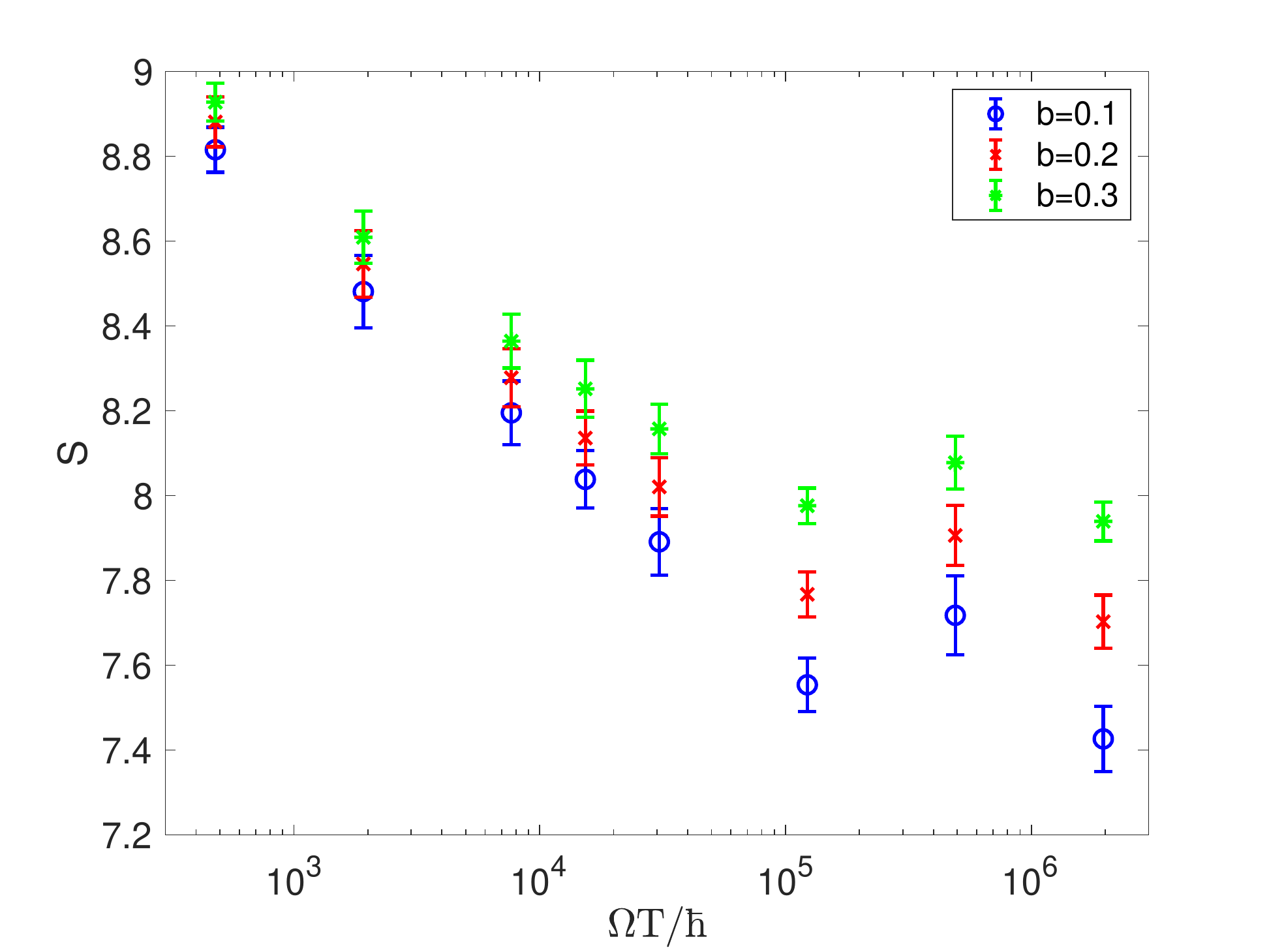}}
\caption{\label{fig:fitvary}
(a) Three alternative versions of the fitted energy distribution, obtained by building up the distribution out of Gaussians with different widths $b$.\\
(b) The three alternative versions of the entropy as a function of sweep time, as in Fig.~\ref{f1}(d) from the main text, obtained using the three alternative fitted energy distributions shown in (a).}
\end{figure}

Which width $b$ was chosen for all the fitting Gaussians had some effect on the fitted distribution, as shown in Fig.~\ref{fig:fitvary}. All reasonable choices of Gaussian width $b$ led to an initially linear decrease of entropy with $\log(T)$ which began to plateau around $\Omega T/\hbar =10^{5}$, as shown in Fig.~\ref{fig:fitvary}(b).

The thermodynamically interesting growth of entropy is inherently related to energy spreading, which is the phenomenon that has been discussed more in the theory of dynamical chaos. The connection between $\Delta E(T)$ and $S(T)$ as functions of sweep time $T$ should hypothetically be direct, because for a probability distribution over energy $P(E,T)$ that is given purely by energy spreading with some $\Delta E(T)$, normalization implies $P(E,T) = f(E/\Delta E)/\Delta E$ for some function $f$ like the ones we see in Fig. S4a. This then implies that
\begin{equation}
S(T) \equiv -\int\!dE\,P(E,T)\log [P(E,T)] = S(T_0) + \log[\Delta E(T)/\Delta E(T_0)] 
\end{equation}
for any reference $T_0$. With $S(T)$ computed from the continuous $P(E,T)$ that we obtained from our finite ensembles by kernel density estimation as described above, our well fitted lines in Fig.~1d and Fig.~S4b show $S(T) = S_0 - a \log (\Omega T/\hbar)$ for $a\sim 0.2$. This implies energy spreading with width that scales with sweep time $T$ as $\Delta E(T) \sim T^{-a}$, that is, with sweep rate as $\Delta E \sim |\dot{x}|^a$, as we note in our main text.

\begin{figure}
\includegraphics[width=0.45 \textwidth]{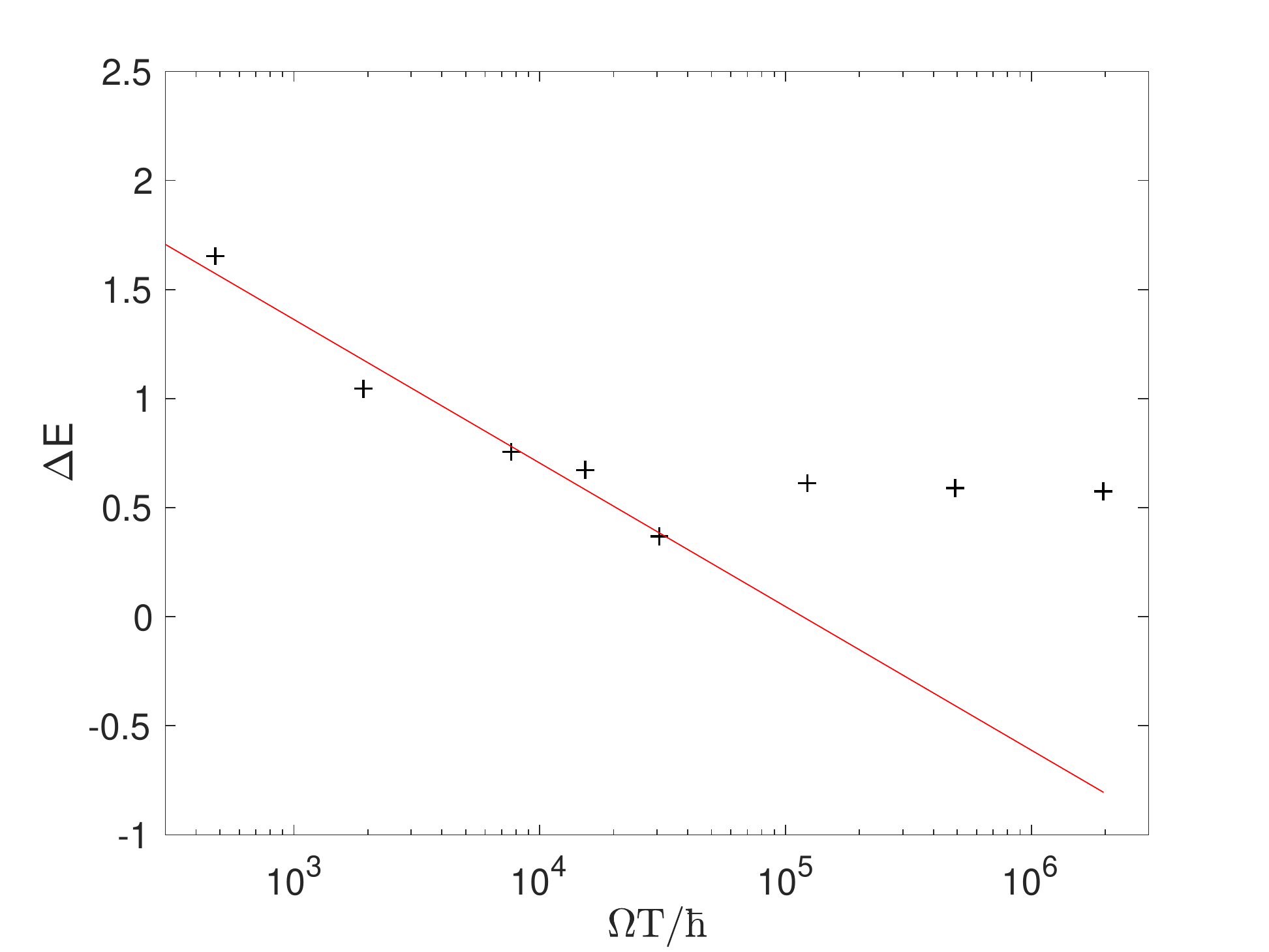}
\caption{\label{fig:DeltaEfit}
Standard deviation of ensemble energy $\Delta E$ at chaos exit time $t_{ex}$ versus sweep duration $T$ as computed directly from our finite ensembles. The red line is a least-squares fit to the first five points; it has a negative slope of approximately $-a = 0.28$.}
\end{figure}

We can also of course directly compute $\Delta E(T)$ as the standard deviation of the energy at $t_{ex}$ in our finite ensembles with various sweep times $T$. The results are shown in Fig.~\ref{fig:DeltaEfit}. This alternative measure of energy spreading shows a somewhat larger fitted slope $a\sim 0.28$ in the power-law regime, albeit with a less good fit than we found from the entropy. It shows an even clearer leveling off of $\Delta E(T)$ to a residual value for $\Omega T/\hbar > 10^5$, with the leveling off perhaps beginning slightly sooner than was clear from the entropy. The gradual re-emergence of an entire ensemble from chaos is such an inherently complex process that even our large ensembles have non-negligible sampling errors; moreover the complexity of the process means that the re-emerged ensemble is not necessarily fully characterized by any single quantity. Even with an infinitely dense ensemble it might well be that the behaviors of $\Delta E(T)$ and $S(T)$ are slightly different. Our conclusion remains clear that, in addition to the known effects of non-adiabatic energy diffusion, the topology of re-emergence from chaos adds a residual energy spreading and entropy production that persists even in the quasi-static limit. Quasi-static irreversibility can occur.

\section{Intermediate values of $u$}
\begin{figure}
\includegraphics[width=0.45 \textwidth]{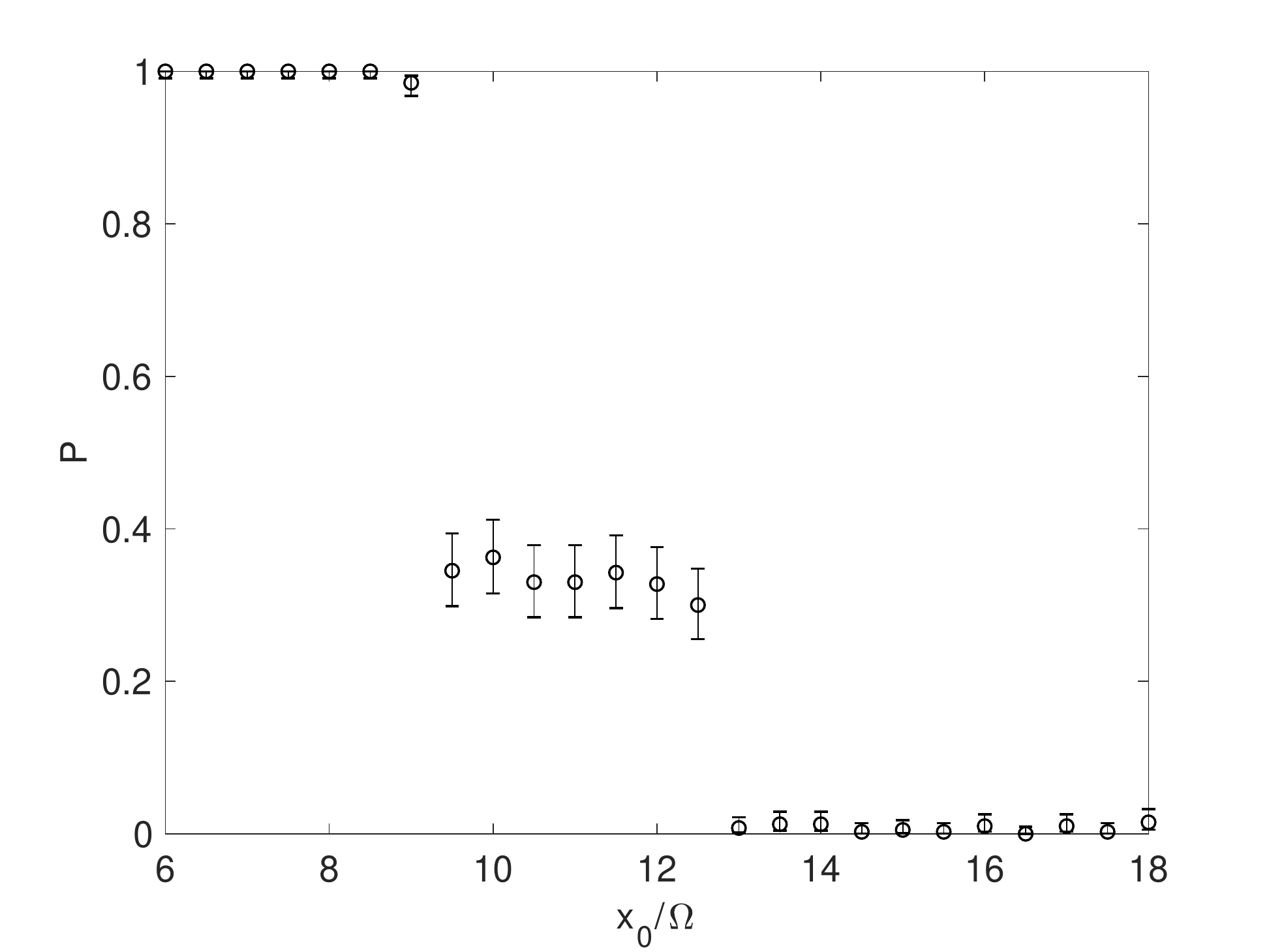}
\caption{\label{fig:integrable_chaotic}
Return probability $P(x_{0})$ for a case with $u=-10$, in which a separatrix crossing and torus merger occur at $x=x_{S}\sim 9$, while longer sweeps that bring $x$ past $x_{C}\sim13$ then bring the ensemble into chaos.}
\end{figure}

In general the two scenarios described above, passage through a separatrix without chaos and passage through chaos, can both happen, one after another, as $x$ is increased. This is the case for a wide range of intermediate values of $u$. The result is that the return probability has two steps: after the separatrix is crossed, but while the motion is still integrable, the return probability has a finite value significantly larger than zero, very similar to the case $u=-40$ discussed above. When the sweep extent is so large that the motion becomes chaotic, the return probability drops from this finite value to almost zero, as in the case $u=-5$. The mechanisms responsible for these two steps are exactly the same as discussed above, namely the merging of tori (first step) and the ergodization on the chaotic part of the energy surface (second step). A simulation for $u=-10$ that demonstrates this behavior is presented in Fig.~\ref{fig:integrable_chaotic}. This simulation also shows the effects of more realistic initial thermal state: instead of the narrow microcanonical distributions assumed in our main text, its initial ensemble is a canonical ensemble with temperature $0.1 \Omega/k_{B}$.

A complication of the hysteresis in this scenario is that portions of the ensemble which have been excited to higher energy during the forward sweep may be chaotic during the backward sweep at $x$ values for which the lower-energy initial ensemble was integrable. There may thus be large portions of the backward sweep during which parts of the ensemble are chaotic while other parts are integrable, in contrast to the relatively brief crossovers between chaos and integrability for the whole ensemble that featured in our main text. Similar behavior to that described in our main text is then seen in each portion of the ensemble separately.

\end{document}